\documentclass[12pt,a4paper,dvips]{article}
\usepackage{a4p}
\usepackage{cite,mcite}
\usepackage{graphicx,rotating}
\usepackage{physics }
\usepackage{epsfig }
\usepackage{l3_title,ifthen}
\usepackage{Lep}

\journalname{Phys. Lett. B}
\date{May 5, 2003}
 \preprint{2003-020}

\usepackage{Lep}
\Lep{2}

\def\Journal#1#2#3#4{{#1} {\bf #2} (#4) #3}

\def\NIMA{{Nucl. Instr. Meth.} A}

\def\PLB{{Phys. Lett.}  B}
\def\PRL{Phys. Rev. Lett.}
\def\PRD{{Phys. Rev.} D}
\def\ZPC{{Z. Phys.} C}
\def\PRP{{Phys. Rep.} C}
\def\CPC{Comput. Phys. Commun.}

\newlength{\capwidth}
\def\pb{\mbox{pb$^{-1}$}}
\newcommand{\EE}{\mathrm{e}^+\mathrm{e}^-}

\newcommand{\gamgam}{\gamma \gamma^*}

\newcommand{\ro}{\mathrm{\rho^0}}
\newcommand{\roro}{\mathrm{\rho^0\rho^0}}
\newcommand{\pipi}{\mathrm{\pi^+\pi^-}}
\newcommand{\q}{ Q^2 }
\newcommand{\mgg}{ W_{\gamma \gamma }}
\newcommand{\ptt}{ p_t^2 }

\begin{document}

\begin{titlepage}
\title{Measurement of Exclusive  $\mathbf{\rho^0\rho^0}$ Production \\
in Two-Photon Collisions  at High $\mathbf{Q^2}$ at LEP}

\author{The L3 Collaboration}


\begin{abstract}

Exclusive $\roro$ production in two-photon 
collisions  involving a single highly virtual photon  
is studied with data collected at LEP 
  at centre-of-mass energies
\mbox{$89 \GeV < \sqrt{s} < 209 \GeV{}$} with a total integrated 
luminosity of $854.7~\pb$. 
The cross section  of the process  $\, \gamma \gamma^* \rightarrow \roro \,$  
is determined as a function of the photon virtuality, $\q$,
and the 
two-photon centre-of-mass energy, $\mgg$,
in the kinematic region:
$1.2 \GeV^2 < \q < 30 \GeV^2$ and 
$1.1 \GeV < W_{\gamma\gamma} < 3 \GeV$. 

\end{abstract}

\submitted

\vfill

\end{titlepage}


\section {Introduction}

The exclusive production of $\ro$ meson pairs in two-photon collisions was studied
by several experiments \cite{TASSO,PLUTO}. 
A prominent feature of the reaction
$\, \gamma \gamma \rightarrow \roro  \,$
is the broad cross section enhancement observed near threshold, 
the origin of which is still not well understood 
 \cite{ATTEMPTS}.
Most   experiments   studied $\roro$ production by quasi-real 
photons, whereas only  scarce data   
involving highly off-shell virtual photons  are available \cite{PLUTO}.
The interest
in  exclusive production of hadron pairs in two-photon 
interactions at high momentum transfer  was recently renewed  since 
methods for calculating the cross section of such processes
were developed 
in the framework of  perturbative QCD \cite{QCD}. In these models, the
exclusive process is factorisable into a perturbative, calculable,
short distance scattering $\rm \gamma\gamma^*\rightarrow q\bar{q}$ or 
 $\rm \gamma\gamma^*\rightarrow gg$ 
and non-perturbative matrix elements describing the transition of the
two partons into hadron pairs, which are called generalized distribution
amplitudes.

This Letter presents results on the study of the two-photon reaction:
\begin{eqnarray}
\label{eq:eqn01}
\EE \to \EE \gamgam \to \EE \roro,
\end{eqnarray}

\noindent
where one of the interacting virtual photons is quasi-real, $\gamma$, and the other one,
 $\gamma ^*$, is highly virtual.
The squared four-momentum, $\q$, of a virtual photon
emitted by the incident beam electron\footnote{Throughout this Letter,   the term ``electron''  denotes
both electrons and positrons.}
is related to the beam energy, $ E_b $, and  to the energy and
scattering angle of the outgoing  electron, $ E_s $ and 
$ \theta_s $ by:
 \begin{eqnarray}
 \label{eq:eqn02}
 \q = 2 E_b E_s (1 - \cos \theta_s)  .
\end{eqnarray}
A scattered electron    detected (``tagged'')
by the forward electromagnetic calorimeter used to measure the luminosity
corresponds to  an off-shell photon with a large  $\q$. 
The rate of such  processes is considerably reduced as compared to production by quasi-real photons
 due to the sharp forward peaking of the angular distribution
of the scattered electron.

The data used in this study  correspond  to an integrated luminosity of 854.7 \pb {} and
were collected by the L3 detector\cite{L3} at LEP.
 Of this sample,  
148.7  \pb~were collected at $\EE$ centre-of-mass energies, \rts , around the Z resonance (Z-pole),
with average  \rts {} of  91 \GeV  {} and 706.0 \pb {} at centre-of-mass energies in the range
$161 \GeV \leq \rts < 209 \GeV$ (high energy), corresponding to an average \rts {} of 195 \GeV .
This Letter presents the production cross section  as a function of $\q$ 
 in the restricted kinematical regions 
\begin{equation}
\label{eq:rangelep1}
1.2 \GeV^2  <  \q  <8.5 \GeV^2 \,\,\,\, \mathrm{(Z-pole)} 
\end{equation}
and
\begin{equation}
\label{eq:rangelep2}
8.8 \GeV^2  <  \q  <30 \GeV^2 \,\,\,\,\,   \mathrm{(high \,\, energy)},
\end{equation}
and  the two-photon mass  interval $1.1 \GeV  <   \mgg   <3 \GeV $. The data
are compared to  Vector Dominance models\cite{GINZBURG} and to a recent QCD model\cite{DIEHLPAP}.



\section {Event Selection}

\subsection {Exclusive four-track events}

The reaction (\ref{eq:eqn01}), contributing to the process
\begin{eqnarray}
\label{eq:eqn03}
\EE \to  \mathrm{e^+ e^-}\pi^+\pi^-\pi^+\pi^- ,
\end{eqnarray}
is identified  by a scattered electron and four charged 
pions measured in the L3 detector.
Tagged two-photon   events are accepted by several independent 
triggers: two   charged-particle triggers  \cite{L3T} and 
an energy trigger demanding a
large energy deposition in the luminosity
monitor in coincidence with at least one track \cite{SEtrig}.
The combined trigger efficiency, as determined
from the data itself, is
$(93.6 \pm 1.3)\%$ at the Z-pole and 
$(97.9 \pm 0.6)\%$ at   high energy.
\par
Single-tagged events are selected by requiring
an electromagnetic cluster
with energy greater then  80\% of the beam energy   reconstructed
in the luminosity monitor, 
 which covers the range   
  $\mathrm{ 25~mrad < \theta < 68~mrad}$ of the electron scattering angle.
   At high energy  the
lower bound increases up to  $\mathrm{ 31~mrad}$ due to the installation of  a mask to protect the
detector from the beam halo.
\par
Event candidates  are required to have exactly four 
tracks, with zero total charge and
with a polar angle, $\theta$, relative to the beam direction, 
such that $\mathrm{ | \cos \theta  |} \le 0.94 $.
A track should come from the interaction vertex and have
transverse momentum greater than  $100 \MeV{}$.
In addition, four-track events, incompatible with the pion mass hypothesis,
are rejected using the 
energy loss information.
\par
Events containing  muons
are removed from the sample.
A search for secondary vertices is performed, and events 
with reconstructed short-lived neutral kaons are rejected.
An event candidate is allowed to contain no more than one electromagnetic cluster, 
with an 
energy below 300 \MeV{} and not exceeding 10\% of the
total energy of the four-pion system.
\par
To ensure that an exclusive final state is detected, 
the  momenta of the tagged electron and the four-pion system should
be well balanced in the plane transverse to the beam direction. Thus
the total transverse momentum squared, $\ptt $, including the scattered electron,
is required to be less than
$ 0.2 \GeV^2$. 
This cut is also effectively a cut on the virtuality of the photon emitted by
the untagged  electron and thus ensures that the $\q$ variable,
calculated from the measured parameters of the tagged electron using (\ref{eq:eqn02}),
corresponds to the photon with the highest virtuality.


\subsection {Background estimation}

\begin{figure} [ht]
  \begin{center}
      \mbox{\epsfig{file=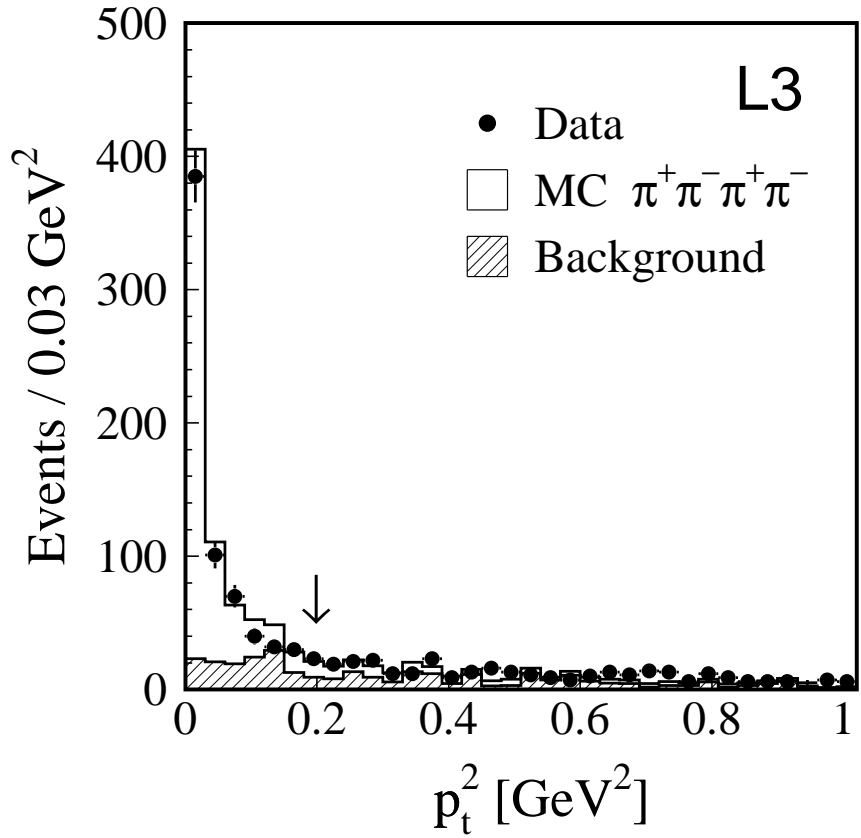,width=0.7\textwidth}}
  \end{center}
  \caption[]{
            The $\ptt$ distribution  
            of the selected $\pipi\pipi$ data (points) 
            in  comparison with   Monte Carlo distributions of four-pion events 
            (open histogram) and the background estimated from the data (hatched histogram). 
            The arrow indicates the selection cut on $\ptt$.}
\label{fig:pt2comp}
\end{figure}

The  contribution to the selected sample due to $\EE$   annihilation   is  negligible.
The background is mainly due to feed-down from tagged two-photon interactions producing 
a higher multiplicity final state, which is incompletely reconstructed. 
To estimate this effect  two background-like data samples are selected.
Firstly, we apply the same selection procedure discussed above
releasing the charge-conservation requirement.
Events of the types $\pi^+ \pi^+ \pi^+ \pi^-$ and 
$\pi^+ \pi^- \pi^- \pi^-$ are selected, in which at least two charged particles were undetected. 
Secondly, we select $\pipi\pipi\pi^0$ events, requiring the $\pipi\pipi$ subsystem  
to pass the four-pion selection discussed above without imposing the $\ptt$ cut,  and  to contain in addition
exactly two  photons with effective mass in the range of $\pm 15 \MeV$ 
around the $\pi^0$ mass. 
We require \mbox{$\ptt  < 0.2 \GeV^2$} in order 
to select tagged exclusive $\pipi\pipi\pi^0$ events, and then consider only their  
$\pipi\pipi$ subsystem  to represent the
background contribution.
We assume that a combination of   these two data samples 
 gives a good description of  the background from partially reconstructed events.
Their $\ptt$ distributions,
  combined with the distribution of reconstructed  
Monte Carlo four-pion events,
agree with the $ \ptt $ distribution observed in the data.
These distributions are shown in  
  Figure~\ref{fig:pt2comp} for the restricted $\q$-ranges (\ref{eq:rangelep1})
 and (\ref{eq:rangelep2}).


\section {Data analysis}

\subsection {Selected Sample}

\begin{figure} [ht]
\mbox{\epsfig{file=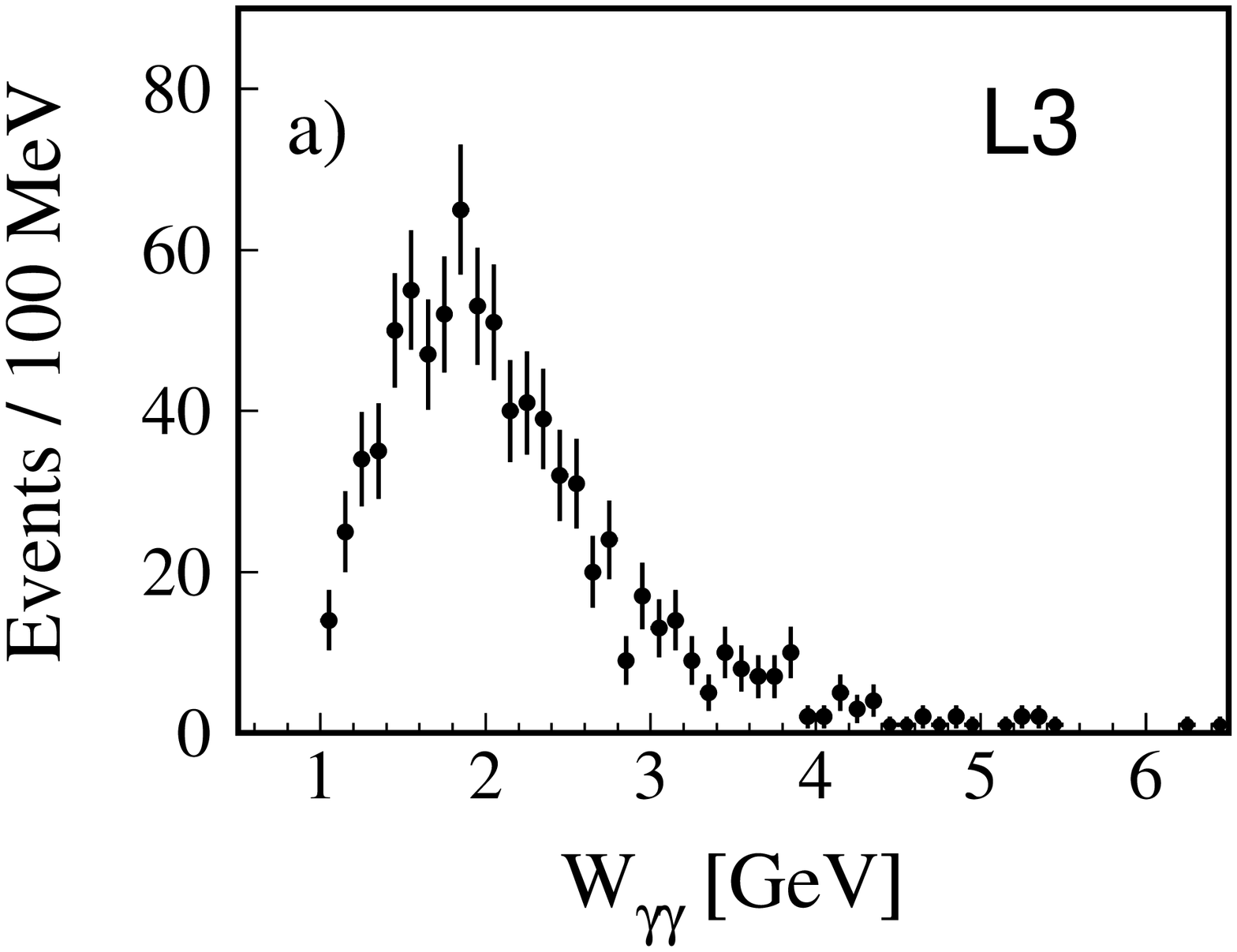,width=0.5\textwidth}}
 \mbox{\epsfig{file=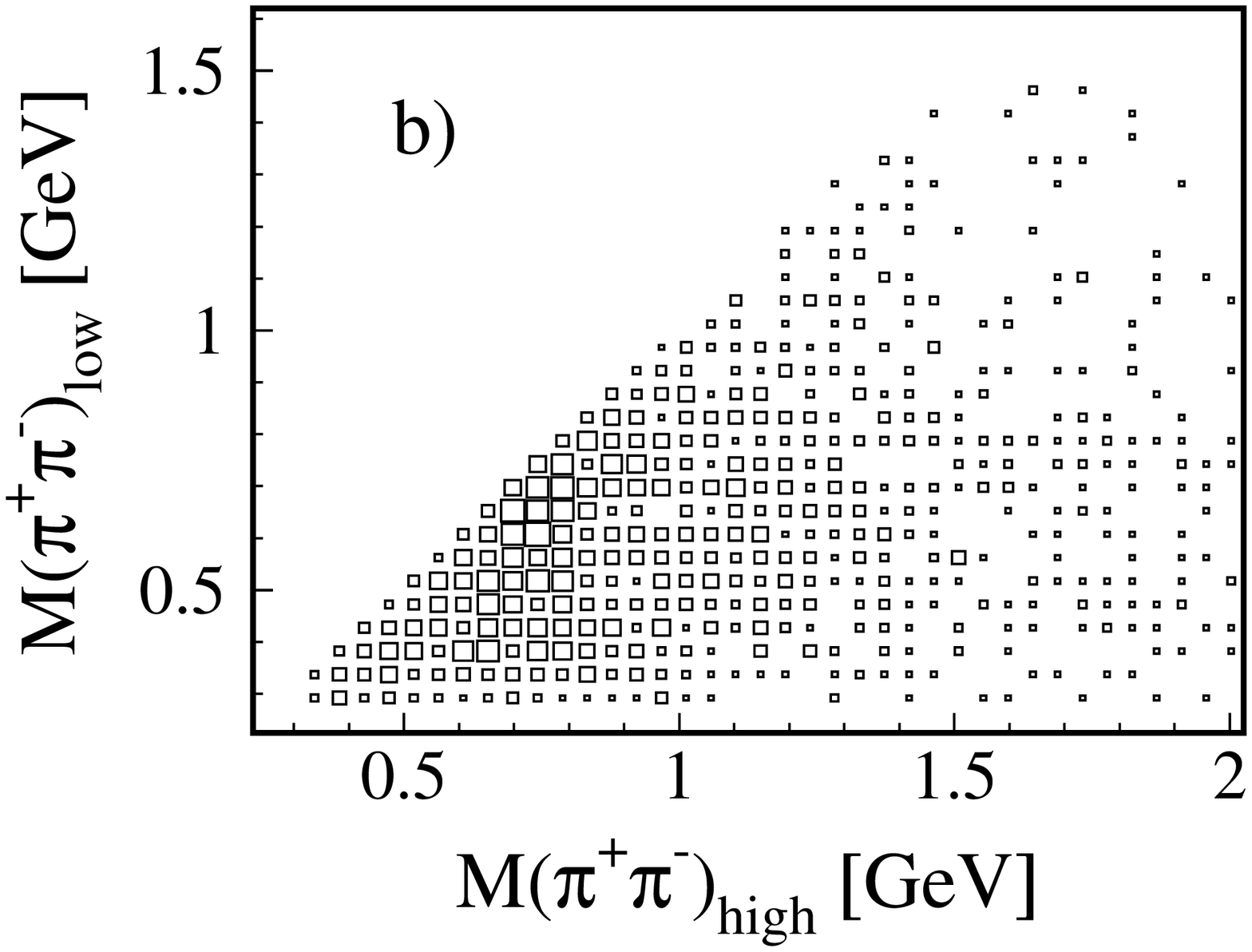,width=0.49\textwidth}}
\mbox{\epsfig{file=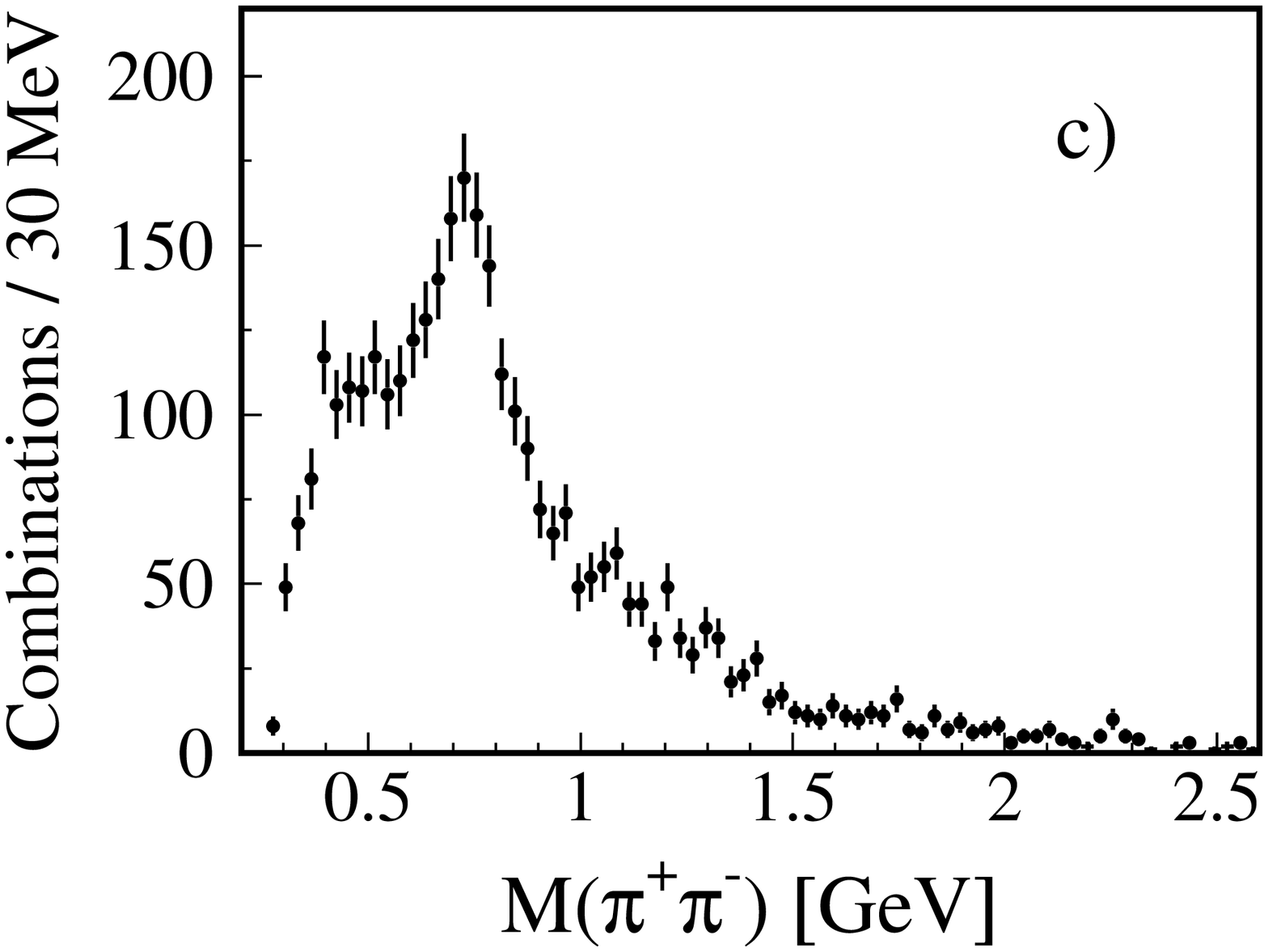,width=0.5\textwidth} }
\mbox{\epsfig{file=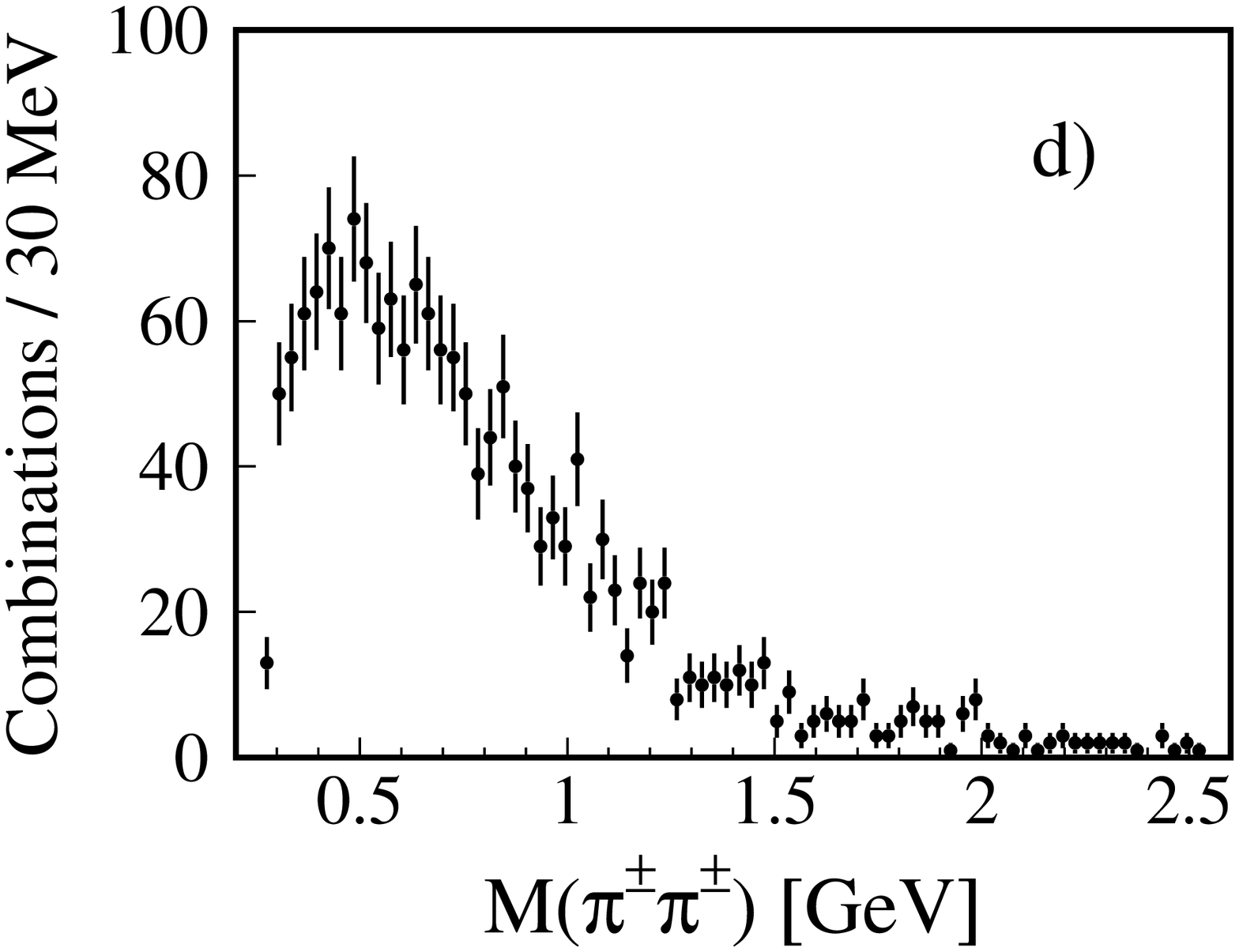,width=0.5\textwidth}}
 
  \caption{Effective mass  distributions for the selected events:
            (a)   Mass of the four-pion system, $ \mgg$.  
            (b) Correlation between the masses of two   $\pipi$ pairs (two entries per event). 
      The     higher mass of each pair is plotted on the horizontal axis.
          (c) Mass of the $\pipi$ combinations (four entries per event). 
         (d) Mass of the $\pi^\pm \pi^\pm$ combinations (two entries per event).           }
\label{fig:rawspec}
\end{figure}

In the region of four-pion   mass $\mgg > 1 \GeV$,
 851 events are selected, 498 events at the Z-pole 
and 353 at high energy. The four-pion mass spectrum of the selected
events is shown in Figure~\ref{fig:rawspec}a.  
\par
The   mass distributions of   $\pipi$ combinations,   
    shown in Figures~\ref{fig:rawspec}b and \ref{fig:rawspec}c, exhibit  a clear $\ro$ signal, while the  mass distribution 
of  $\pi^\pm \pi^\pm$ combinations, shown 
in Figure~\ref{fig:rawspec}d, has no resonance structure. In Figure~\ref{fig:rawspec}b,
 the clustering of entries  in the region of the crossing of the $\rho^0$ mass-bands 
 gives evidence for a contribution 
 of $\roro$ intermediate states.
\par
The $\roro$ production rate is determined as a function of $\q$ and $\mgg$. 
The resolution of the reconstructed variables $\q$  and  $\mgg$ is
  better than 3\%   and thus  the  
event migration between adjacent   bins is negligible.


\subsection {Monte Carlo  Modelling}

To estimate the number of $\roro$ events in the selected four-pion data sample,
we consider non-interfering contributions from three processes: 
\begin{eqnarray}
\label{eq:eqn04}
&&  \gamgam \to \roro;    \nonumber \\
&&  \gamgam \to \ro \pipi;  \\
&&  \gamgam \to \pipi \pipi ,\, \mathrm{non-resonant.} \nonumber  
\end{eqnarray}
\par
The data statistics is not sufficient to reach conclusions about 
contributions from subprocesses involving production of higher-mass resonances 
such as the $f_2(1270)$.  Therefore in the present analysis we assume 
that the data is  described by the processes (\ref{eq:eqn04}) only.
It was demonstrated that such a model provides a good description of 
exclusive four-pion production by quasi-real photons \cite{TASSO}.
\par 
 Monte Carlo samples of the process (\ref{eq:eqn04}) are generated
 with the  EGPC~\cite{LINDE} program. About two million events of each process
 are produced for both the  Z-pole and the high energy regions.
The $\mgg$ and $\q$-dependence  are those of the  
$\gamma\gamma$ luminosity function~\cite{BUDNEV} 
and  only isotropic 
production and phase space decays are included.
These events 
are  processed in the same way as the data, 
 introducing specific detector inefficiencies   for the different data taking periods.


For acceptance calculations, the Monte Carlo events are assigned a $\q$-dependent weight,
evaluated using the  GVDM~\cite{GVDM} form-factor for both photons.
Taking  
 into account the detector acceptance and the efficiency of the 
selection procedure, the detection efficiency for each $\q$  and  $\mgg$ bin
  is listed in Tables~\ref{tbl:xsectq2}--\ref{tbl:xsectwgg_lep2}.  It
  is in the range of $10\% - 25\%$, almost independent of the process. It
  slowly increases with $\q$
and slowly decreases with $\mgg$.


\subsection {Fit Method}

In order to determine the differential 
$\roro$ production rate, a maximum likelihood fit to the data of the sum of the processes (\ref{eq:eqn04})
is performed in intervals of $\q$ and $\mgg$.
The set, $\Omega$, of six two-pion masses, 
 the four   $\pipi$ combinations and the  two $\pi^\pm \pi^\pm$ combinations, 
provides a complete description of a four-pion event in our model of isotropic
production and decay discussed above.
This choice of kinematic variables allows to fully exploit the information 
specific to each one of the processes  (\ref{eq:eqn04}) and to obtain their 
contributions to the observed four-pion yield.
For each data event, $i$, with measured variables $\Omega_i$, we calculate the
probabilities, 
$\mathrm{P}_j(\Omega_i)$, that the event resulted
from   the production mechanism  $j$.
The likelihood function is defined as:
\begin{eqnarray}
\label{eq:eqn05}
&& \Lambda = \prod_{i}   \sum_{j=1}^{3} \lambda_j \mathrm{P}_j (\Omega_i) 
 \,,\;\;\;\;\;\; \;\;\sum_{j=1}^{3} \lambda_j  = 1,
\end{eqnarray}
where $\lambda_j$ is the fraction of the process $j$ in the
$\pipi\pipi$ sample for a given  $\q$ or $\mgg$  bin
and  the product runs over all data events in that bin.
 The probabilities $\mathrm{P}_j$ are determined  by the 
six-fold differential cross sections of the corresponding process,
using Monte Carlo   samples and a box method\cite{BOXMETHOD}.   

\par
The fitting procedure is tested by applying it on various mixtures of Monte Carlo
event samples from the processes (\ref{eq:eqn04}), treated as data. 
The  contribution  of the $\roro$ production process is always reproduced  within 
statistical uncertainties, whereas,
 for small statistics test samples,
large negative correlations, in the range of 60\% -- 75\%,
exist between the $\ro \pipi$ and $\pipi\pipi$ (non-resonant)
  fractions. 
Both contributions are, however, necessary to fit the data. Therefore, in the following, 
 only the $\ro\ro$ content and the sum of the  $\ro \pipi$ and $\pipi\pipi$ (non-resonant)
contributions are considered.
\par
To check the quality of the fit, 
the two-pion mass
distributions of the data are compared with those of a mixture of Monte Carlo event samples 
from the processes (\ref{eq:eqn04}),  in the proportion  determined by the fit.
The data and Monte Carlo distributions 
 are in   good agreement
over the whole $\q$ and $\mgg$ range;
 an example is shown in  Figure~\ref{fig:composq2}.
\par
\begin{figure} [p]
  \begin{center}
    \vskip -2cm
    \mbox{\epsfig{file=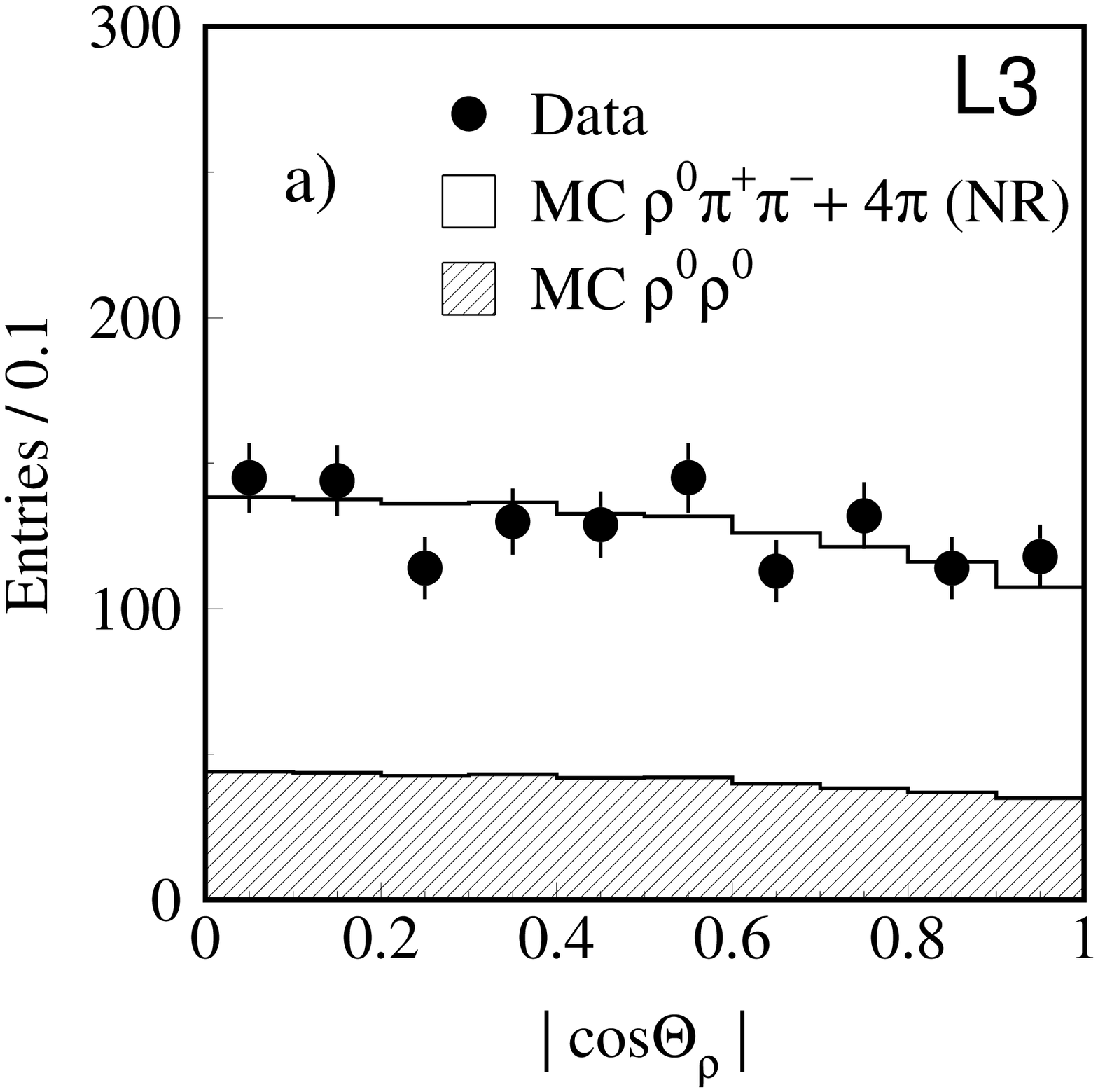,width=0.49\textwidth}}
    \mbox{\epsfig{file=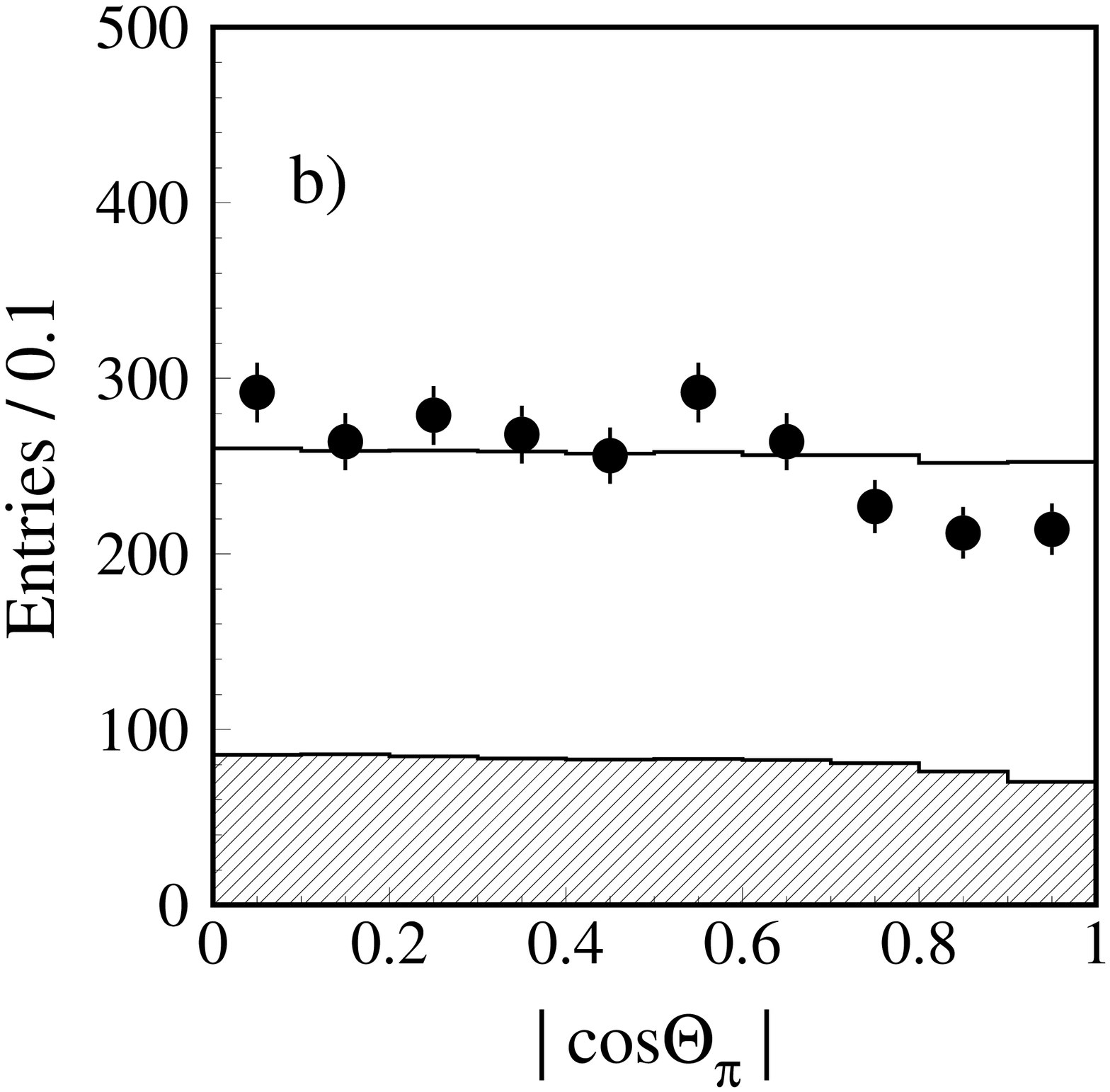,,width=0.49\textwidth}}

    \mbox{\epsfig{file=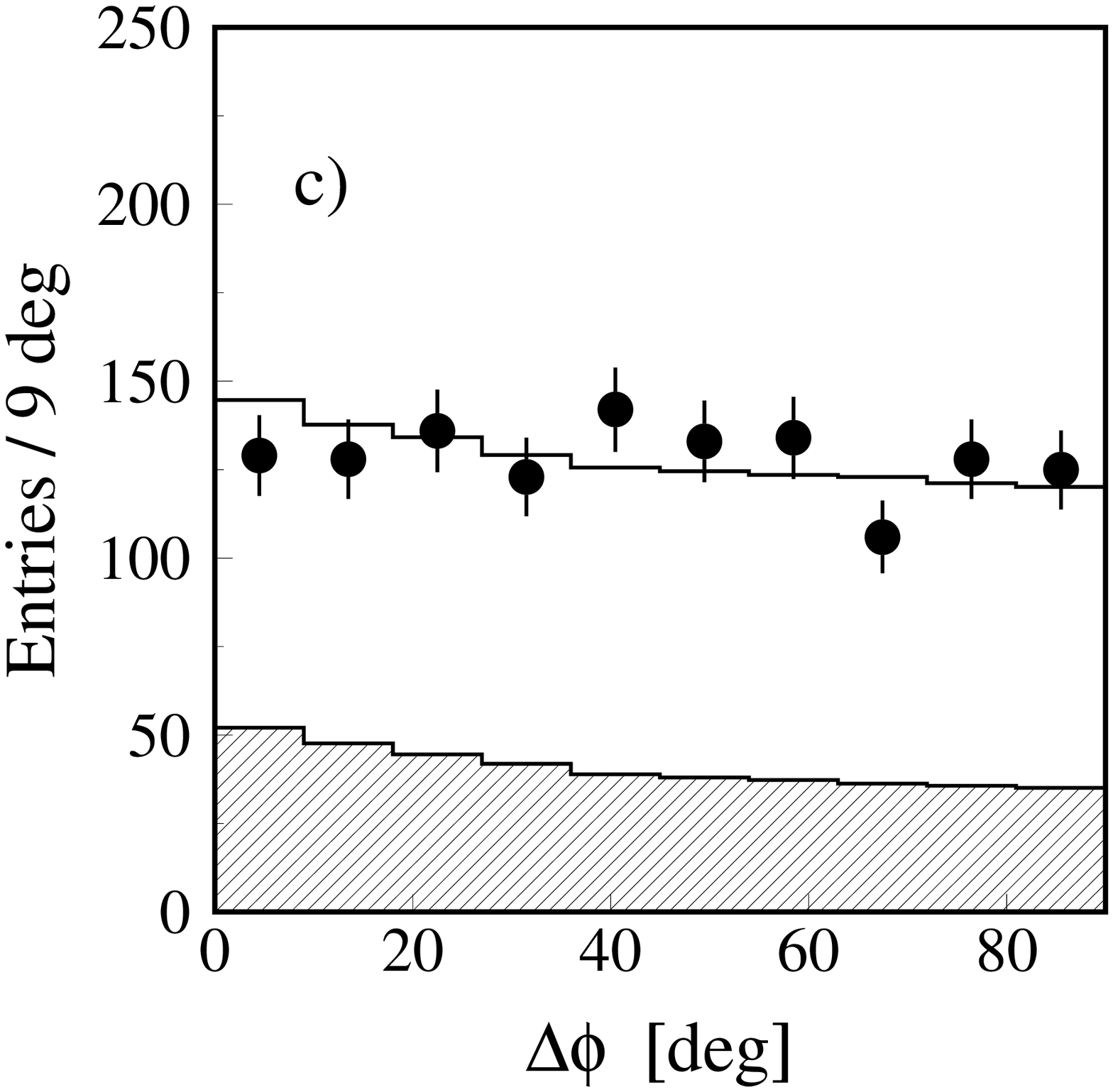,,width=0.49\textwidth}}
    \mbox{\epsfig{file=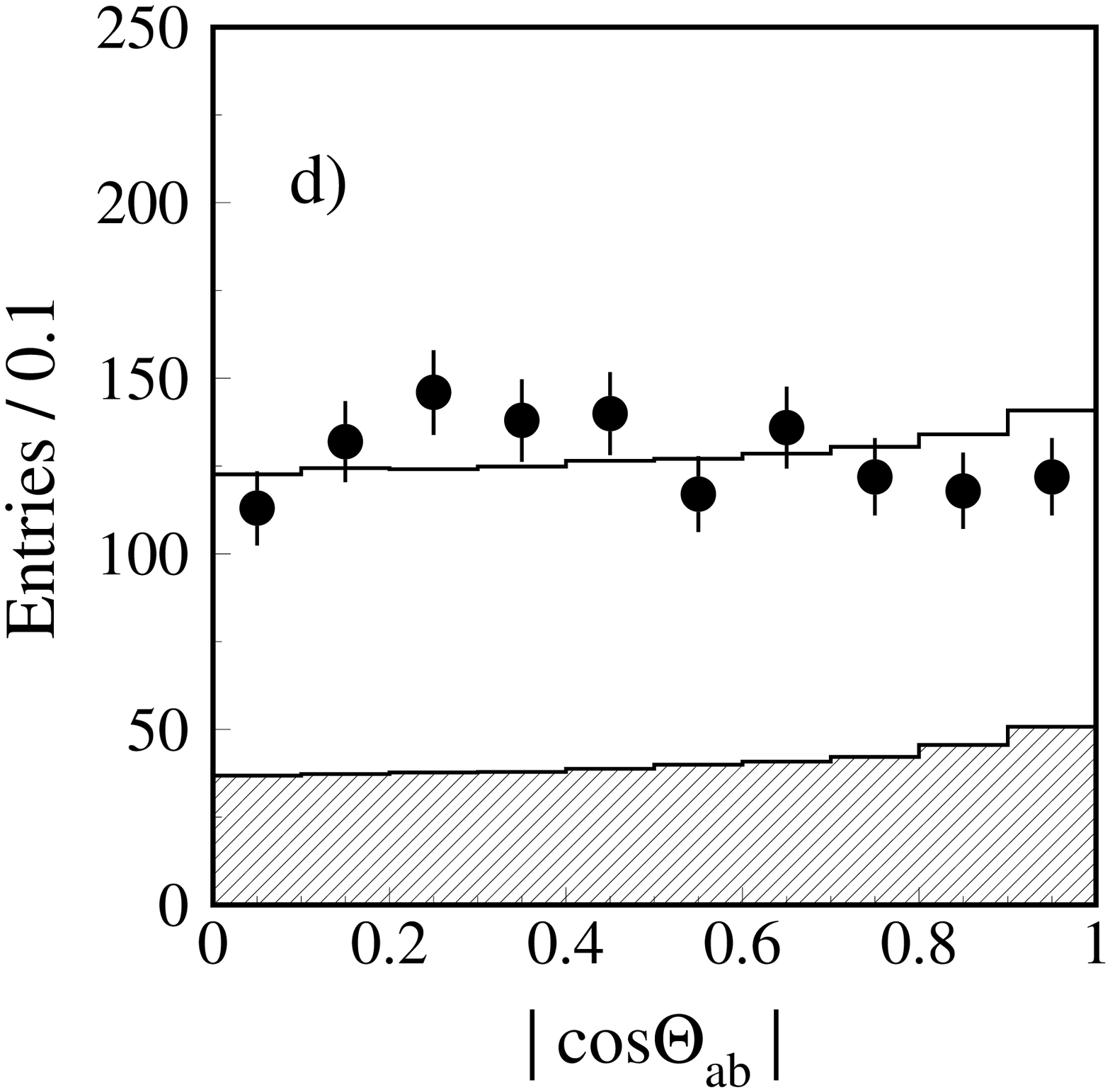,,width=0.49\textwidth}}

  \end{center}
  \caption[]{Comparison of data   and   Monte Carlo   angular distributions:
           (a) $\mid \cos \theta_\rho \mid$,  the cosine of the polar angle of 
           the $\ro$ with respect to the $\gamma\gamma$ axis in the $\gamma\gamma$ 
           center-of-mass system; 
           (b) $\mid \cos \theta_\pi \mid$, the cosine of the polar angle of the pion 
           in its parent $\ro$ helicity-system; 
           (c) $\Delta \phi$, the angle between the decay planes of the two $\ro$ mesons
           in the $\gamma\gamma$ centre-of-mass system;
           (d) $\mid \cos \theta_\mathrm{ab} \mid$, the cosine of the opening angle between 
           the two $\pi^+$ directions of flight, each one defined in its parent $\ro$ 
           rest-system.   
           There are two entries per event in (a),(c) and (d) and four entries per event in (b).
           The points represent the data, the hatched area shows  the $\ro\ro$ component 
	   and the open area shows the sum of
             $\ro \pipi$ and $\pipi\pipi$ (non-resonant) components. 
           The fraction of the different components are determined by the fit and the
	  total normalisation is to the number of the events.
           }
\label{fig:angles}
\end{figure}

 
As   pointed out in Reference \citen{ACHASOV}, the $\pipi$ system in the  $\ro \pipi$ final state 
cannot have an isotropic angular distribution, since, in order to conserve C-parity, 
the angular momentum between the two pions has to 
be odd.
We have verified that our results are   insensitive to variations
of the underlying angular distributions in the production model. In addition,
a good agreement of the measured angular distributions of the data
with those of the Monte Carlo is observed,  as  presented in Figure~\ref{fig:angles}.



\section {Results}

\subsection {Cross Sections}

The  cross sections, $\mathrm{\Delta \sigma_{ee}}$, of the process $\EE \to   \EE \roro $
are measured as a function of  $\q$ and $\mgg$ and  
 are listed in Tables~\ref{tbl:xsectq2}--\ref{tbl:xsectwgg_lep2}  together with the 
 efficiencies and background contamination. 
The statistical uncertainties, listed in the Tables, follow
from the fit.  
The differential cross section $d \sigma_{\mathrm{ee}} / d \q$ of the process 
(\ref{eq:eqn01}), derived from $\mathrm{\Delta \sigma_{ee}}$, 
is also listed in Table~\ref{tbl:xsectq2}.
When evaluating the differential cross section,
a correction,  based on the $\q$-dependence of the $\roro$ Monte Carlo sample,
is applied, such as to assign the cross section value to the centre 
of the corresponding $\q$-bin~\cite{BIN}.
 
\par
 To evaluate the cross section  $\sigma_{\gamma\gamma}$ of the process
$ \gamgam \to \ro \ro $, the integral of the transverse photon luminosity function, 
$L_{TT}$, 
is computed for each $\q$ and $\mgg$  bin using the   program GALUGA \cite{GALUGA}, 
which performs exact QED calculations. The cross section $\sigma_{\gamma\gamma}$
is derived from  the measured cross section 
$\mathrm{\Delta \sigma_{ee}}$
using the relation
$\mathrm{\Delta \sigma_{ee}} = L_{TT}  \sigma_{\gamma\gamma}$. 
Thus  $\sigma_{\gamma\gamma}$ represents
an effective cross section containing contributions from both transverse ($T$) and
longitudinal ($L$) photon polarizations:
\begin{eqnarray}
\label{eq:eqn09}
\sigma_{\gamma\gamma}(\mgg,\q) = \sigma_{TT}(\mgg,\q) + \epsilon   \sigma_{TL}(\mgg,\q)   \,,
\end{eqnarray}
where $\sigma_{TT} $ and $\sigma_{TL} $ are the cross sections for collision 
of transverse-transverse and transverse-longitudinal photons.
The ratio of longitudinal to transverse polarization of the virtual photon, $\epsilon$,
given, approximately, by the expression:
\begin{eqnarray}
\label{eq:eqn10}
 \epsilon \approx \frac{ 2  E_{s} / E_{b}}   {1+(E_{s}/E_{b})^2}  \, ,
\end{eqnarray}
is greater than 0.98, for our data.
\par
\begin{figure} [p]
  \begin{center}
    \vskip -2cm
    \mbox{\epsfig{file=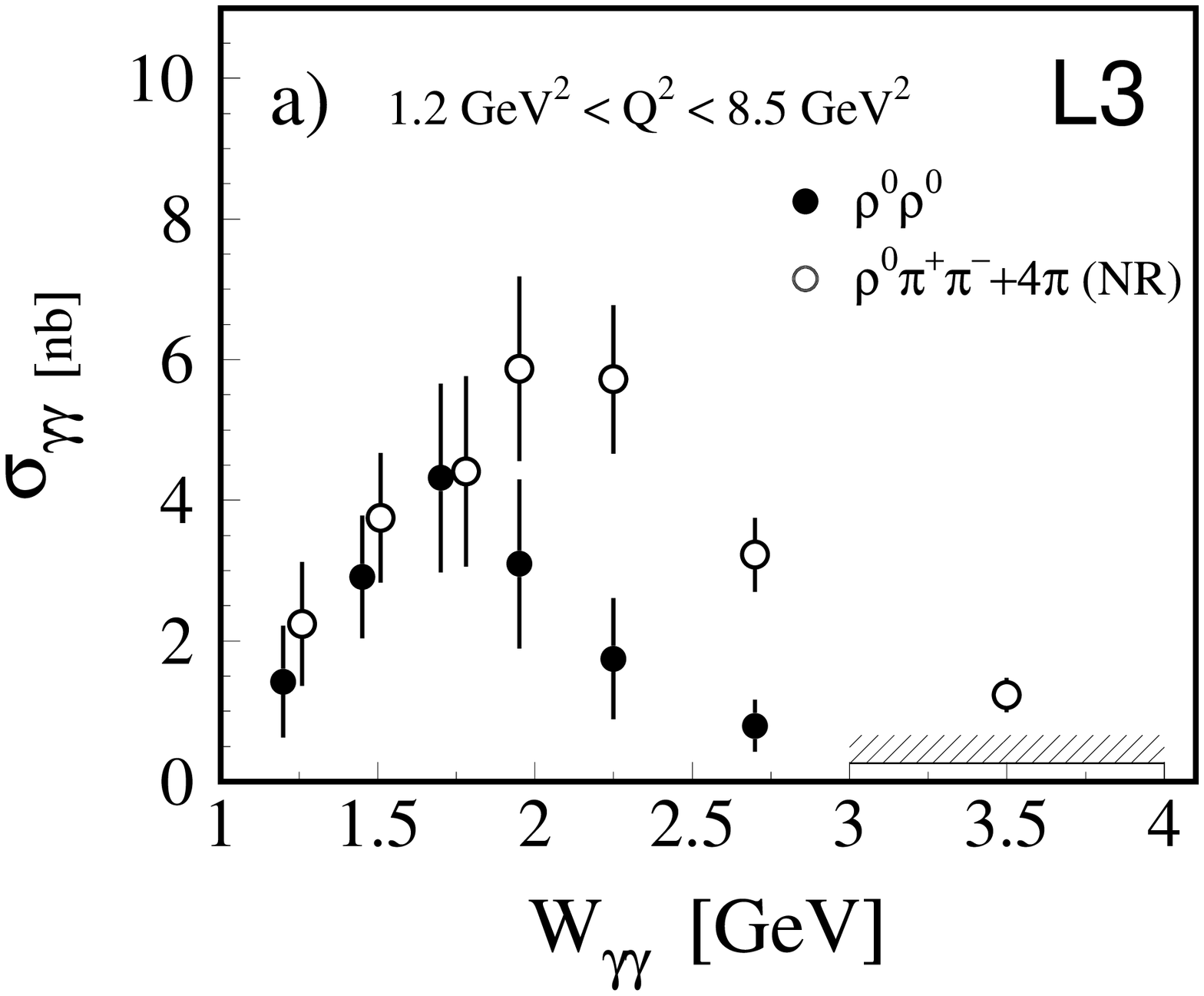,width=0.8\textwidth}}
    \vskip -1cm
    \mbox{\epsfig{file=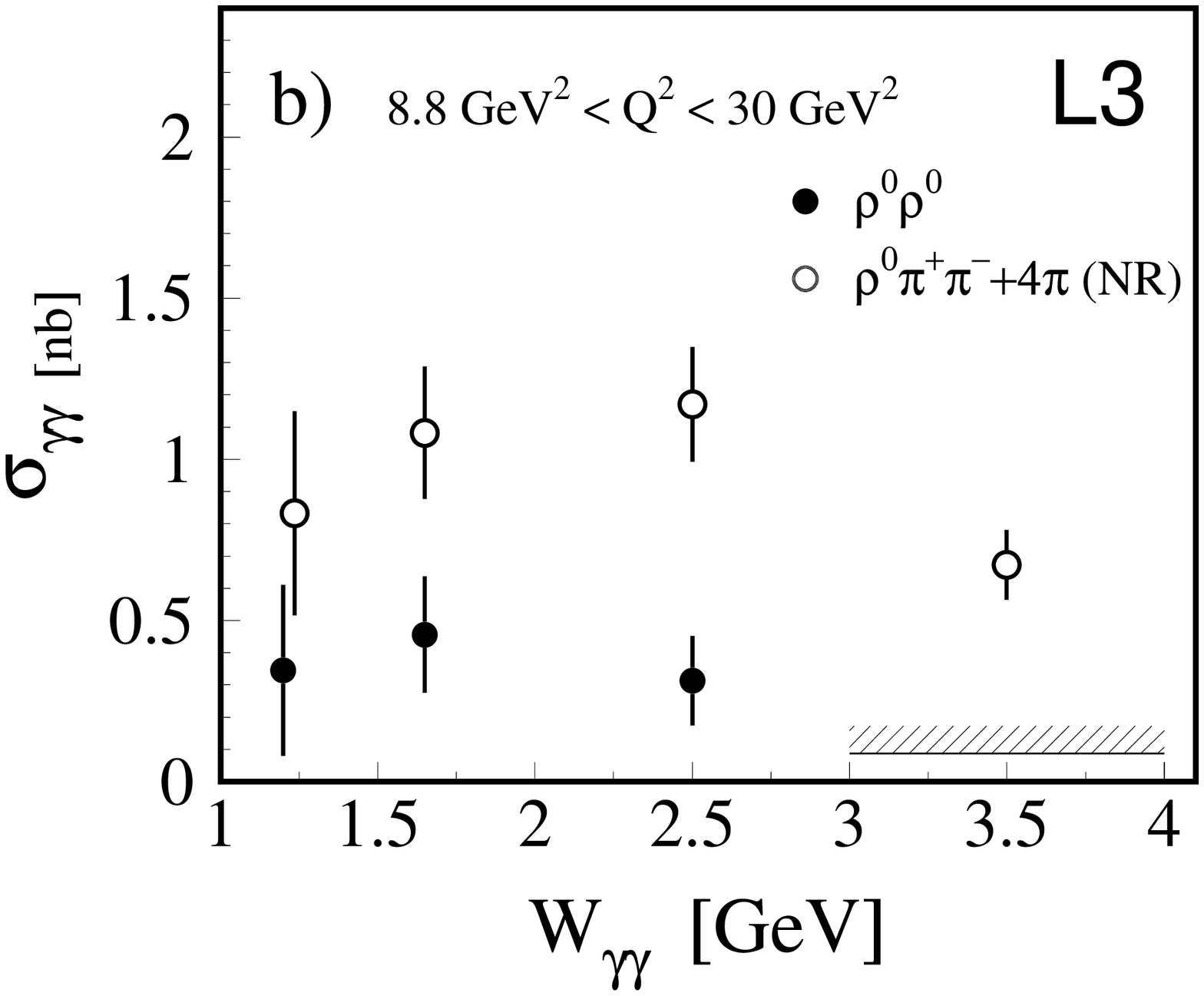,width=0.8\textwidth}}
  \end{center}
  \caption[]{
            Cross section of the process $\gamgam\to\ro\ro$
            and the sum of the cross sections of the processes
            $\gamgam \to \ro\pipi$
            and $\gamgam \to \pipi\pipi$(non-resonant) 
            as  functions of  $\mgg$,  for 
            (a) $1.2 \GeV^2 < \q < 8.5 \GeV^2$ and 
            (b) $8.8 \GeV^2 < \q < 30 \GeV^2$.
            The points represent the data, the  bars show the 
            statistical uncertainties. The horizontal line for
	    the highest $\mgg$ bin indicates the upper limit of the $\gamgam\to\ro\ro$ 
	    cross section at 95\% CL: 0.26 nb for the Z-pole data and
	    0.087 nb for the high energy data.
	    
           }
\label{fig:xsectwgg}
\end{figure}

The cross section of the process $ \gamgam \to \ro \ro $ as a function of $\mgg$,
 listed in Table~\ref{tbl:xsectwgg_lep1} and \ref{tbl:xsectwgg_lep2},
is plotted in Figure~\ref{fig:xsectwgg} 
together with the  sum of the cross sections of the processes
$ \gamgam \to  \ro\pipi$ 
and $ \gamgam \to \pipi\pipi$(non-resonant). 
The statistical uncertainties of the sum of these two cross sections  take into account their 
correlations.
The $ \ro \ro $ cross section is dominated
by a broad   enhancement at threshold,
 already observed in the data at 
$\q \approx 0$\cite{TASSO} and at moderate $\q$
\cite{PLUTO}.
The two cross sections are  
 listed in Table \ref{tbl:xsectq2} and plotted in
 Figure~\ref{fig:xsectq2}a  as a function of $\q$ .

\subsection {Systematics}

The uncertainty on this measurement is dominated by statistics. 
The uncertainty on the measured cross section due to  the selection procedure,  estimated 
by varying the cuts,  is in the range  $7\% - 20\%$, 
affecting more the higher $\q$ region.
Different form-factor expressions  used for reweighting the Monte Carlo events 
and the variation of the acceptance
contribute to an overall shift in the range  $ 2\% - 6\% $. 
The fitting procedure uncertainty mostly
depends  on the box size.
It is estimated to be in the
range $7\% - 18\% $ for the fits in $\q$ and   in the range $8\% - 30\% $
for the fits in   $\mgg$.

To estimate the uncertainties of the background correction, 
the background determination procedure is performed using only the
$\pi^\pm \pi^\pm \pi^\pm \pi^\mp$ 
or only the  $\pipi\pipi\pi^0$ samples. 
A contribution in the range  $6\% -11\%$ is obtained.
\par
Collinear initial state radiation has little impact on the measurement 
since for 91\% of the selected events the energy of the tagged electron exceeds 
90\% of the beam energy.
\par
All the  contributions are
  added in quadrature to obtain the  
   systematic uncertainties quoted in   Tables~\ref{tbl:xsectq2}--\ref{tbl:xsectwgg_lep2}.

\subsection {Fits to the Data}

\begin{figure} [ht]
  \begin{center}
    \vskip -1.5cm
    \mbox{\epsfig{file=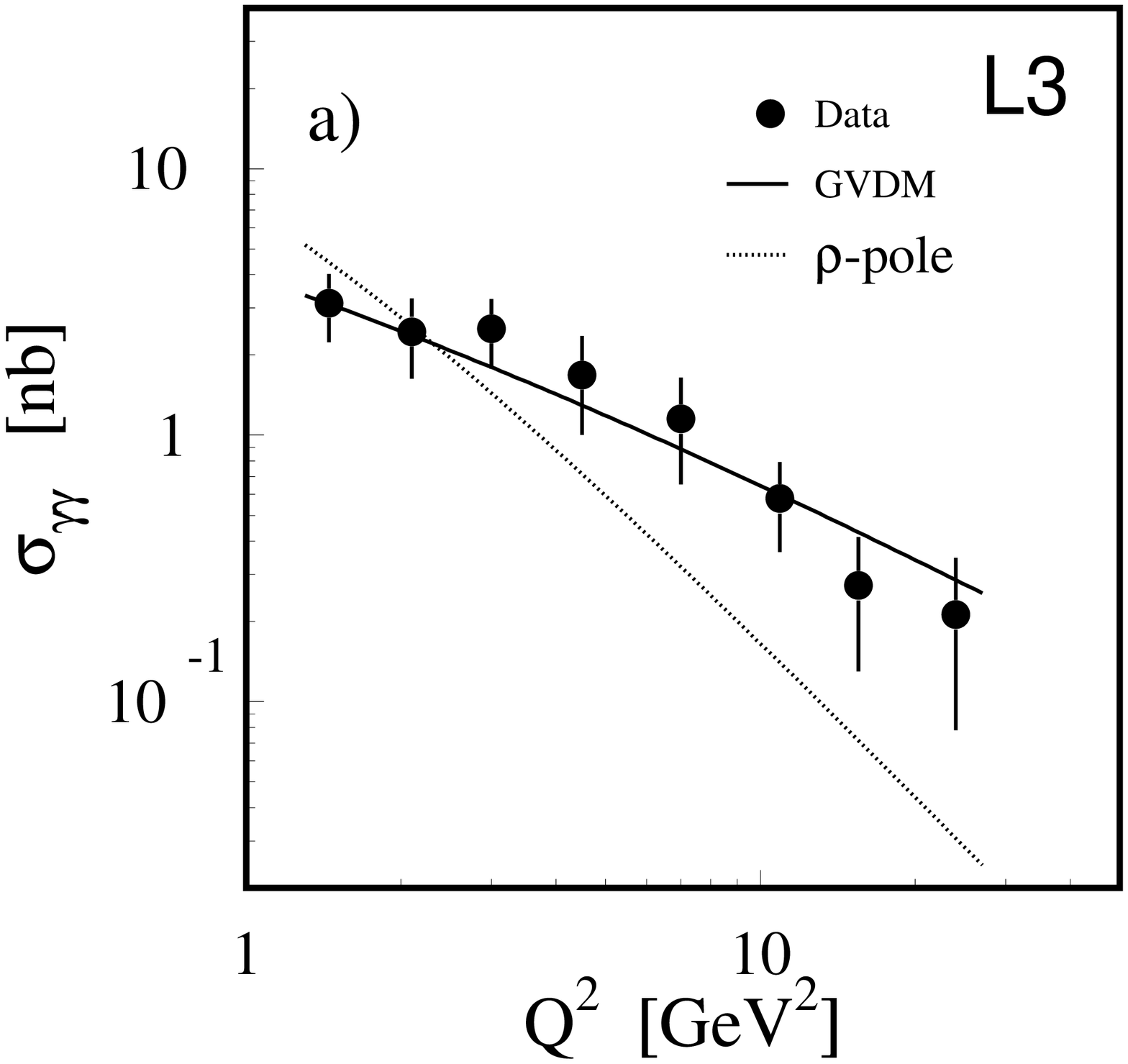,width=0.65\textwidth}}
    \vskip -1cm
    \mbox{\epsfig{file=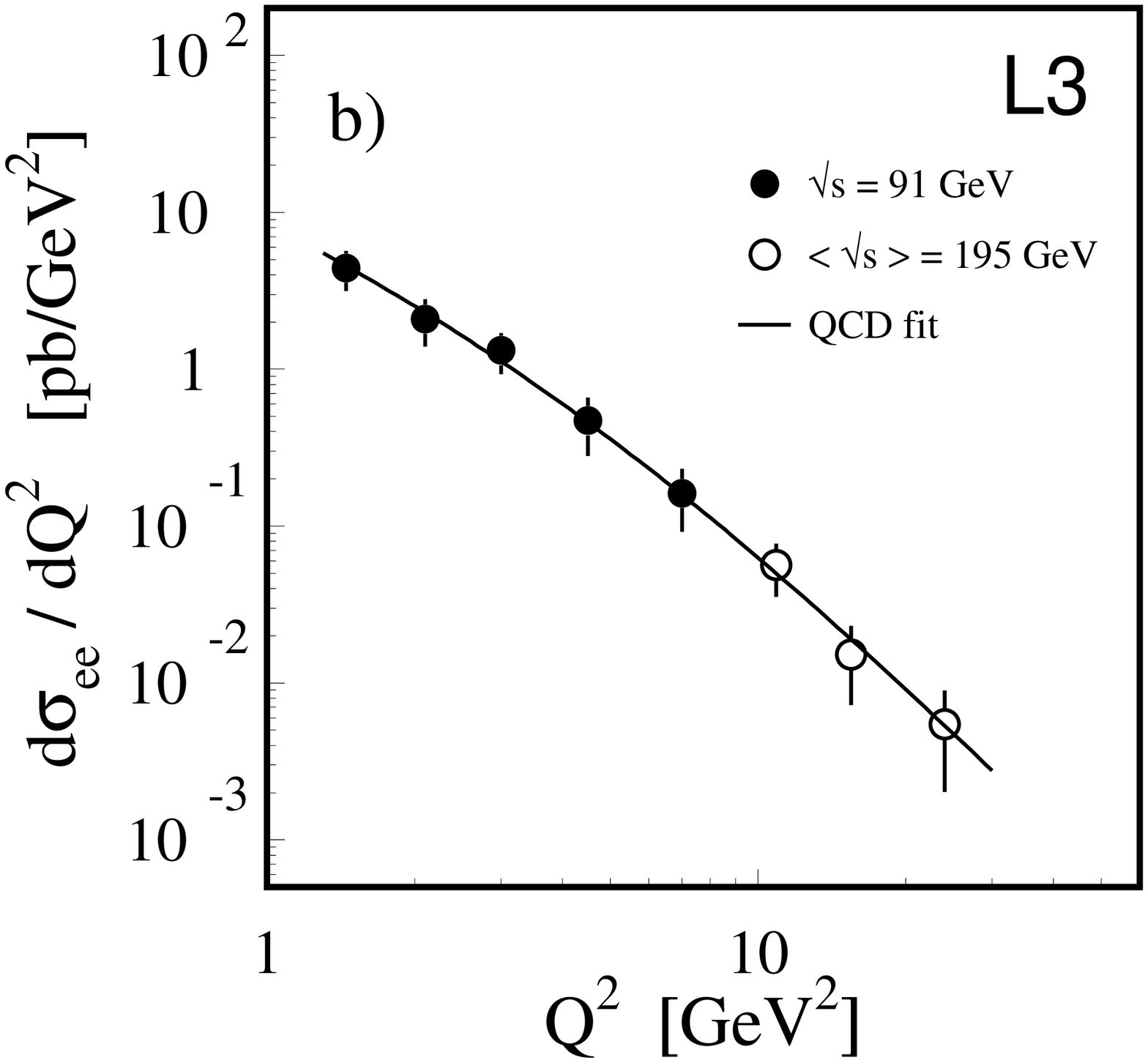,width=0.65\textwidth}}
  \vspace{-3mm}
  \end{center}
 \caption{The $\roro$ production cross section as a function of 
         $\q$, for $1.1 \GeV < \mgg < 3 \GeV$:
         (a) cross section of the process $\gamgam \to \ro \ro$ and 
         (b) differential cross section of the process 
         $ \EE \to \EE \roro$.
	 The points represent the data, the  bars show the
         statistical uncertainties.
         The solid line in (a) represents the result of a fit based on 
         the generalized vector dominance model \protect\cite{GINZBURG} 
         and the dotted line indicates the expectation for  a $\rho$-pole form-factor.
         The line in (b) represents the result of a fit to a form
         expected from QCD   calculations.
           }
\label{fig:xsectq2}
\end{figure}

Figure~\ref{fig:xsectq2}b shows the result of
a fit of the differential cross section 
$d \sigma_{\mathrm{ee}} / d \q$ to a form~\cite{DIEHL} 
expected from QCD-based calculations\cite{DIEHLPAP}:
\begin{eqnarray}
\label{eq:eqn11}
 d \sigma_{\mathrm{ee}} / d \q \sim 
\frac{1} { Q^n (\q + < \mgg >^2)^2} \,.
\end{eqnarray}
The fit  is performed using the central value of the mass spectrum    $< \mgg > = 1.94 \GeV $. It provides 
a good description of the $\q$-dependence of the data with
  an exponent $ n = 2.4 \pm 0.3$, to be compared with the expected value 
$ n = 2$. Only statistical uncertainties are considered.
A common fit of the data   taken at the Z-pole and at high energy
is justified by the almost constant values
of the photon polarization parameter $\epsilon$, which determines
the energy dependence of the cross section. 
\par
In Figure~\ref{fig:xsectq2}a the data are fitted with two different form-factor parametrisations,
leaving  the normalization as a free parameter.
A form
suggested in Reference \citen{GINZBURG},    based on the generalized vector dominance model
(GVDM) \cite{GVDM}, 
  provides a good description of the $\q$-dependence of the data
  whereas a steeper decrease 
is expected for a simple $\rho$-pole form-factor.

\section*{Acknowledgements}

We thank M. Diehl and O. Teryaev for very useful discussions.


%
%

\newpage
\typeout{   }     
\typeout{Using author list for paper 261 -  }
\typeout{$Modified: Jul 15 2001 by smele $}
\typeout{!!!!  This should only be used with document option a4p!!!!}
\typeout{   }
%
%
%
%
%
%

\newcount\tutecount  \tutecount=0
\def\tutenum#1{\global\advance\tutecount by 1 \xdef#1{\the\tutecount}}
\def\tute#1{$^{#1}$}
\tutenum\aachen            
\tutenum\nikhef            
\tutenum\mich              
\tutenum\lapp              
\tutenum\basel             
\tutenum\lsu               
\tutenum\beijing           
\tutenum\bologna           
\tutenum\tata              
\tutenum\ne                
\tutenum\bucharest         
\tutenum\budapest          
\tutenum\mit               
\tutenum\panjab            
\tutenum\debrecen          
\tutenum\dublin            
\tutenum\florence          
\tutenum\cern              
\tutenum\wl                
\tutenum\geneva            
\tutenum\hefei             
\tutenum\lausanne          
\tutenum\lyon              
\tutenum\madrid            
\tutenum\florida           
\tutenum\milan             
\tutenum\moscow            
\tutenum\naples            
\tutenum\cyprus            
\tutenum\nymegen           
\tutenum\caltech           
\tutenum\perugia           
\tutenum\peters            
\tutenum\cmu               
\tutenum\potenza           
\tutenum\prince            
\tutenum\riverside         
\tutenum\rome              
\tutenum\salerno           
\tutenum\ucsd              
\tutenum\sofia             
\tutenum\korea             
\tutenum\purdue            
\tutenum\psinst            
\tutenum\zeuthen           
\tutenum\eth               
\tutenum\hamburg           
\tutenum\taiwan            
\tutenum\tsinghua          

{
\parskip=0pt
\noindent
{\bf The L3 Collaboration:}
\ifx\selectfont\undefined
 \baselineskip=10.8pt
 \baselineskip\baselinestretch\baselineskip
 \normalbaselineskip\baselineskip
 \ixpt
\else
 \fontsize{9}{10.8pt}\selectfont
\fi
\medskip
\tolerance=10000
\hbadness=5000
\raggedright
\hsize=162truemm\hoffset=0mm
\def\r{\rlap,}
\noindent

P.Achard\r\tute\geneva\ 
O.Adriani\r\tute{\florence}\ 
M.Aguilar-Benitez\r\tute\madrid\ 
J.Alcaraz\r\tute{\madrid}\ 
G.Alemanni\r\tute\lausanne\
J.Allaby\r\tute\cern\
A.Aloisio\r\tute\naples\ 
M.G.Alviggi\r\tute\naples\
H.Anderhub\r\tute\eth\ 
V.P.Andreev\r\tute{\lsu,\peters}\
% ...
\newpage
\begin{list}{A}{\itemsep=0pt plus 0pt minus 0pt\parsep=0pt plus 0pt minus 0pt
                \topsep=0pt plus 0pt minus 0pt}
\item[\aachen]
 III. Physikalisches Institut, RWTH, D-52056 Aachen, Germany$^{\S}$
\item[\nikhef] National Institute for High Energy Physics, NIKHEF, 
     and University of Amsterdam, NL-1009 DB Amsterdam, The Netherlands
\item[\mich] University of Michigan, Ann Arbor, MI 48109, USA
\item[\lapp] Laboratoire d'Annecy-le-Vieux de Physique des Particules, 
     LAPP,IN2P3-CNRS, BP 110, F-74941 Annecy-le-Vieux CEDEX, France
\item[\basel] Institute of Physics, University of Basel, CH-4056 Basel,
     Switzerland
\item[\lsu] Louisiana State University, Baton Rouge, LA 70803, USA
\item[\beijing] Institute of High Energy Physics, IHEP, 
  100039 Beijing, China$^{\triangle}$ 
\item[\bologna] University of Bologna and INFN-Sezione di Bologna, 
     I-40126 Bologna, Italy
\item[\tata] Tata Institute of Fundamental Research, Mumbai (Bombay) 400 005, India
\item[\ne] Northeastern University, Boston, MA 02115, USA
\item[\bucharest] Institute of Atomic Physics and University of Bucharest,
     R-76900 Bucharest, Romania
\item[\budapest] Central Research Institute for Physics of the 
     Hungarian Academy of Sciences, H-1525 Budapest 114, Hungary$^{\ddag}$
\item[\mit] Massachusetts Institute of Technology, Cambridge, MA 02139, USA
\item[\panjab] Panjab University, Chandigarh 160 014, India.
\item[\debrecen] KLTE-ATOMKI, H-4010 Debrecen, Hungary$^\P$
\item[\dublin] Department of Experimental Physics,
  University College Dublin, Belfield, Dublin 4, Ireland
\item[\florence] INFN Sezione di Firenze and University of Florence, 
     I-50125 Florence, Italy
\item[\cern] European Laboratory for Particle Physics, CERN, 
     CH-1211 Geneva 23, Switzerland
\item[\wl] World Laboratory, FBLJA  Project, CH-1211 Geneva 23, Switzerland
\item[\geneva] University of Geneva, CH-1211 Geneva 4, Switzerland
\item[\hefei] Chinese University of Science and Technology, USTC,
      Hefei, Anhui 230 029, China$^{\triangle}$
\item[\lausanne] University of Lausanne, CH-1015 Lausanne, Switzerland
\item[\lyon] Institut de Physique Nucl\'eaire de Lyon, 
     IN2P3-CNRS,Universit\'e Claude Bernard, 
     F-69622 Villeurbanne, France
\item[\madrid] Centro de Investigaciones Energ{\'e}ticas, 
     Medioambientales y Tecnol\'ogicas, CIEMAT, E-28040 Madrid,
     Spain${\flat}$ 
\item[\florida] Florida Institute of Technology, Melbourne, FL 32901, USA
\item[\milan] INFN-Sezione di Milano, I-20133 Milan, Italy
\item[\moscow] Institute of Theoretical and Experimental Physics, ITEP, 
     Moscow, Russia
\item[\naples] INFN-Sezione di Napoli and University of Naples, 
     I-80125 Naples, Italy
\item[\cyprus] Department of Physics, University of Cyprus,
     Nicosia, Cyprus
\item[\nymegen] University of Nijmegen and NIKHEF, 
     NL-6525 ED Nijmegen, The Netherlands
\item[\caltech] California Institute of Technology, Pasadena, CA 91125, USA
\item[\perugia] INFN-Sezione di Perugia and Universit\`a Degli 
     Studi di Perugia, I-06100 Perugia, Italy   
\item[\peters] Nuclear Physics Institute, St. Petersburg, Russia
\item[\cmu] Carnegie Mellon University, Pittsburgh, PA 15213, USA
\item[\potenza] INFN-Sezione di Napoli and University of Potenza, 
     I-85100 Potenza, Italy
\item[\prince] Princeton University, Princeton, NJ 08544, USA
\item[\riverside] University of Californa, Riverside, CA 92521, USA
\item[\rome] INFN-Sezione di Roma and University of Rome, ``La Sapienza",
     I-00185 Rome, Italy
\item[\salerno] University and INFN, Salerno, I-84100 Salerno, Italy
\item[\ucsd] University of California, San Diego, CA 92093, USA
\item[\sofia] Bulgarian Academy of Sciences, Central Lab.~of 
     Mechatronics and Instrumentation, BU-1113 Sofia, Bulgaria
\item[\korea]  The Center for High Energy Physics, 
     Kyungpook National University, 702-701 Taegu, Republic of Korea
\item[\purdue] Purdue University, West Lafayette, IN 47907, USA
\item[\psinst] Paul Scherrer Institut, PSI, CH-5232 Villigen, Switzerland
\item[\zeuthen] DESY, D-15738 Zeuthen, Germany
\item[\eth] Eidgen\"ossische Technische Hochschule, ETH Z\"urich,
     CH-8093 Z\"urich, Switzerland
\item[\hamburg] University of Hamburg, D-22761 Hamburg, Germany
\item[\taiwan] National Central University, Chung-Li, Taiwan, China
\item[\tsinghua] Department of Physics, National Tsing Hua University,
      Taiwan, China
\item[\S]  Supported by the German Bundesministerium 
        f\"ur Bildung, Wissenschaft, Forschung und Technologie
\item[\ddag] Supported by the Hungarian OTKA fund under contract
numbers T019181, F023259 and T037350.
\item[\P] Also supported by the Hungarian OTKA fund under contract
  number T026178.
\item[$\flat$] Supported also by the Comisi\'on Interministerial de Ciencia y 
        Tecnolog{\'\i}a.
\item[$\sharp$] Also supported by CONICET and Universidad Nacional de La Plata,
        CC 67, 1900 La Plata, Argentina.
\item[$\triangle$] Supported by the National Natural Science
  Foundation of China.
\end{list}
}
\vfill


\newpage
\newpage

\begin{table}[htb]
\begin{center}
\begin{sideways}
\begin{minipage}[b]{\textheight}
\begin{center}
\begin{tabular}{|c|c|c|l|l|c|l|}
\hline
$ \q$-range &  $\varepsilon$ & $\mathit{Bg}$ & $~~~~~~\Delta \sigma_{ee}$ [ pb ] & $ \;\, d\,\sigma_{ee}/d\,\q$ [ pb\,/$\GeV^2 \,]$ & 
$ \sigma_{\gamma\gamma}$ [ nb ]  & $ \, d\,\sigma_{ee}/d\,\q$ [ pb\,/$\GeV^2 \,]$ \\
 $[\;\GeV \;\;]$ & [ \% ] & [ \% ] & $~~~~~~~~~~~~\roro$ & $~~~~~~~~~~~~~~\roro$ &  $\roro$ &  $ ~~\ro\pipi + \pipi\pipi$  \\ \hline

\phantom{0}1.2 -- \phantom{0}1.7  & \phantom{0}9   & 31         & $   2.30\phantom{0} \pm 0.66\phantom{0}  \pm 0.28\phantom{0} $ & $ 4.41\phantom{00} \pm 1.26\phantom{00} \pm 0.53\phantom{00} $  & $ 3.13 \pm 0.89 \pm 0.38 $ & $ 6.77\phantom{0} \pm 1.35\phantom{00} \pm 0.87\phantom{00} $ \\ \hline
\phantom{0}1.7 -- \phantom{0}2.5  & 12             & 27         & $   1.76\phantom{0} \pm 0.59\phantom{0}  \pm 0.26\phantom{0} $ & $ 2.09\phantom{00} \pm 0.70\phantom{00} \pm 0.31\phantom{00} $  & $ 2.44 \pm 0.81 \pm 0.37 $ & $ 3.82\phantom{0} \pm 0.76\phantom{00} \pm 0.49\phantom{00} $ \\ \hline
\phantom{0}2.5 -- \phantom{0}3.5  & 14             & 21         & $   1.36\phantom{0} \pm 0.40\phantom{0}  \pm 0.25\phantom{0} $ & $ 1.32\phantom{00} \pm 0.39\phantom{00} \pm 0.24\phantom{00} $  & $ 2.51 \pm 0.74 \pm 0.45 $ & $ 1.74\phantom{0} \pm 0.41\phantom{00} \pm 0.22\phantom{00} $ \\ \hline
\phantom{0}3.5 -- \phantom{0}5.5  & 15             & 16         & $   1.02\phantom{0} \pm 0.41\phantom{0}  \pm 0.17\phantom{0} $ & $ 0.47\phantom{00} \pm 0.19\phantom{00} \pm 0.080\phantom{0} $  & $ 1.68 \pm 0.68 \pm 0.29 $ & $ 0.97\phantom{0} \pm 0.21\phantom{00} \pm 0.16\phantom{00} $ \\ \hline
\phantom{0}5.5 -- \phantom{0}8.5  & 18             & \phantom{0}6 & $ 0.53\phantom{0} \pm 0.23\phantom{0}  \pm 0.10\phantom{0} $ & $ 0.16\phantom{00} \pm 0.070\phantom{0} \pm 0.030\phantom{0} $  & $ 1.15 \pm 0.50 \pm 0.22 $ & $ 0.33\phantom{0} \pm 0.083\phantom{0} \pm 0.046\phantom{0} $ \\ \hline
\phantom{0}8.8 -- 13.0            & 16             & 11           & $ 0.25\phantom{0} \pm 0.094            \pm 0.038 $           & $ 0.056\phantom{0} \pm 0.021\phantom{0} \pm 0.0085           $  & $ 0.58 \pm 0.21 \pm 0.09 $ & $ 0.17\phantom{0} \pm 0.025\phantom{0} \pm 0.019\phantom{0} $ \\ \hline
13.0 -- 18.0                      & 21             & 23           & $ 0.080           \pm 0.042            \pm 0.022$            & $ 0.015\phantom{0} \pm 0.0079           \pm 0.0041           $  & $ 0.27 \pm 0.14 \pm 0.07 $ & $ 0.044           \pm 0.0096           \pm 0.0065           $ \\ \hline
18.0 -- 30.0                      & 22             & 23           & $ 0.075           \pm 0.047            \pm 0.022$            & $ 0.0055           \pm 0.0035           \pm 0.0017           $  & $ 0.21 \pm 0.13 \pm 0.06 $ & $ 0.013           \pm 0.0040           \pm 0.0027           $ \\ \hline
  
\end{tabular}
\end{center}

\caption{Detection efficiencies, $\varepsilon$,  background fractions, $\mathit{Bg}$,  
         and measured production cross sections
         as a function of  $\q$ for $1.1 \GeV < \mgg < 3 \GeV$
	 for   Z-pole and high energy data.
         The values of the differential cross sections are   corrected to the centre of each bin.
         }
\label{tbl:xsectq2}
\end{minipage}
\end{sideways}
\end{center}
\end{table}

\begin{table*}[ht]
\begin{center}
\begin{tabular}{|c|c|c|c|c|c|}
\hline
$  \mgg$-range & $\varepsilon$&  $\mathit{Bg}$ &  $ \Delta \sigma_{ee}$ [ pb ] & 
$ \sigma_{\gamma\gamma}$ [ nb ]  & $ \sigma_{\gamma\gamma}$ [ nb ] \\
$[\;\GeV \;\;]$ & [ \% ] & [ \% ] &  $\roro$ &  $\roro$ &  $ \ro\pipi + \pipi\pipi$  \\ \hline
1.1 -- 1.3    & 11 & 18 & $ 0.56 \pm 0.31  \pm 0.14  $  & $ 1.42 \pm 0.79 \pm 0.36 $ & $ 2.24 \pm 0.88 \pm 0.41 $ \\ \hline
1.3 -- 1.6    & 11 & 26 & $ 1.64 \pm 0.49  \pm 0.28  $  & $ 2.91 \pm 0.87 \pm 0.50 $ & $ 3.75 \pm 0.92 \pm 0.56 $ \\ \hline
1.6 -- 1.8    & 12 & 25 & $ 1.53 \pm 0.47  \pm 0.26  $  & $ 4.32 \pm 1.34 \pm 0.73 $ & $ 4.41 \pm 1.35 \pm 0.79 $ \\ \hline
1.8 -- 2.1    & 13 & 23 & $ 1.54 \pm 0.60  \pm 0.35  $  & $ 3.10 \pm 1.20 \pm 0.71 $ & $ 5.87 \pm 1.31 \pm 0.88 $ \\ \hline
2.1 -- 2.4    & 13 & 22 & $ 0.80 \pm 0.39  \pm 0.20  $  & $ 1.75 \pm 0.86 \pm 0.44 $ & $ 5.72 \pm 1.06 \pm 0.81 $ \\ \hline
2.4 -- 3.0    & 13 & 20 & $ 0.64 \pm 0.30  \pm 0.13  $  & $ 0.79 \pm 0.37 \pm 0.16 $ & $ 3.22 \pm 0.52 \pm 0.39 $ \\ \hline
3.0 -- 4.0    & 14 & 15 & $ 0.09 \pm 0.12  \pm 0.07  $  & $ 0.08 \pm 0.11 \pm 0.06 $ & $ 1.23 \pm 0.24 \pm 0.17 $ \\ \hline
\end{tabular}
\caption{Detection efficiencies, $\varepsilon$,  background fractions, $\mathit{Bg}$,
         and measured production cross sections 
         as a function of  $\mgg$, for
         $1.2 \GeV^2 < \q < 8.5 \GeV^2$, for  the Z-pole data.
         The cross sections of the reaction
         $ \EE \to \EE \roro$, $\gamgam \to \roro$ 
	and  of the sum of the processes $\gamgam \to \ro\pipi$ and $\gamgam \to \pipi\pipi$(non-resonant) are also given. 
         The first uncertainties are statistical, the second systematic.
         }
\label{tbl:xsectwgg_lep1}
\end{center}
\end{table*}


\begin{table*}[ht]
\begin{center}
\begin{tabular}{|c|c|c|c|c|c|c|}
\hline
$ \mgg$-range & $\varepsilon$ &  $\mathit{Bg}$ &  $ \Delta \sigma_{ee}$ [ pb ] & 
$   \sigma_{\gamma\gamma}$ [ nb ]  & $\sigma_{\gamma\gamma}$ [ nb ] \\
$[\;\GeV \;\;]$& [ \% ] & [ \% ] & $\roro$  & $\roro$ & $ \ro\pipi + \pipi\pipi$ \\ \hline
1.1 -- 1.3    & 22  & 19 & $ 0.041 \pm 0.032 \pm 0.017 $   & $ 0.34 \pm 0.26 \pm 0.14 $ & $ 0.83 \pm 0.32 \pm 0.27 $ \\ \hline
1.3 -- 2.0    & 19  & 23 & $ 0.184 \pm 0.073 \pm 0.046 $    & $ 0.46 \pm 0.18 \pm 0.11 $ & $ 1.08 \pm 0.21 \pm 0.15 $ \\ \hline
2.0 -- 3.0    & 17  & 9 & $ 0.175 \pm 0.077 \pm 0.053 $   & $ 0.31 \pm 0.14 \pm 0.09 $ & $ 1.17 \pm 0.18 \pm 0.16 $ \\ \hline
3.0 -- 4.0    & 16  & 9 & $ 0.012 \pm 0.022 \pm 0.009 $   & $ 0.02 \pm 0.04 \pm 0.02 $ & $ 0.67 \pm 0.11 \pm 0.09 $ \\ \hline
\end{tabular}
\caption{Detection efficiency, $\varepsilon$,  background fractions, $\mathit{Bg}$,
         and measured production cross sections
         as a function of  $\mgg$, for
         $8.8 \GeV^2 < \q < 30  \GeV^2$, for the high energy data.
         The cross section of the reaction
         $ \EE \to \EE \roro$,  $\gamgam \to \roro$
         and  of the sum of the processes
         $\gamgam \to \ro\pipi$ and $\gamgam \to \pipi\pipi$(non-resonant) are also given. 
         The first uncertainties are statistical, the second systematic.
         }
\label{tbl:xsectwgg_lep2}
\end{center}
\end{table*}

%
%
%
\clearpage
\vfil

\clearpage
\vfil


\hskip -0.5cm
\begin{minipage}{.5cm}
\begin{center}
\large 
\rotatebox{90}{Entries / 80 MeV}\\
\normalsize
\end{center}
\end{minipage}\begin{minipage}{16.5cm}

\vskip -1.5cm

{\epsfig{file=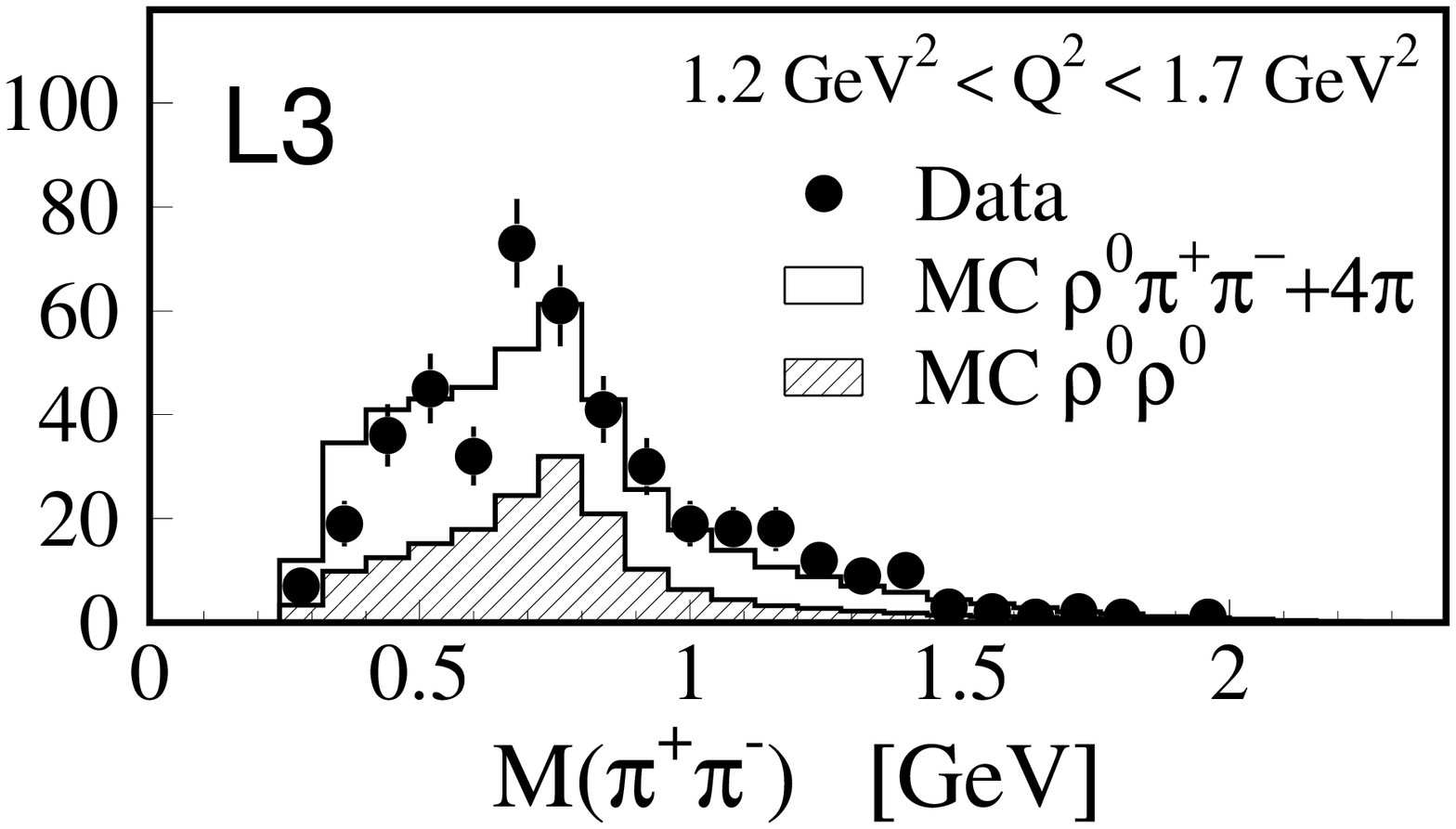,width=0.49\linewidth}}
{\epsfig{file=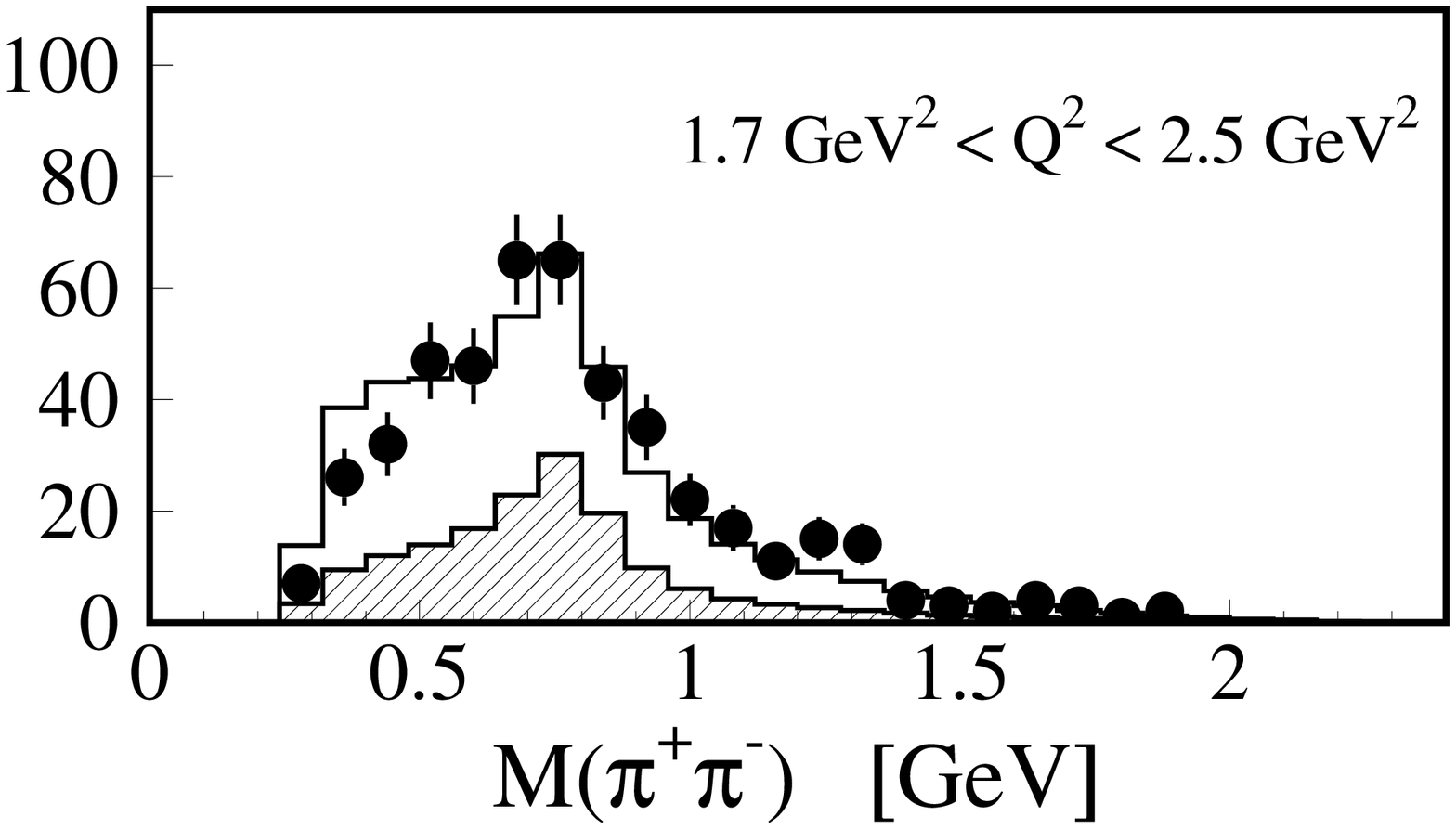,width=0.49\linewidth}}
\vskip -0.5cm
{\epsfig{file=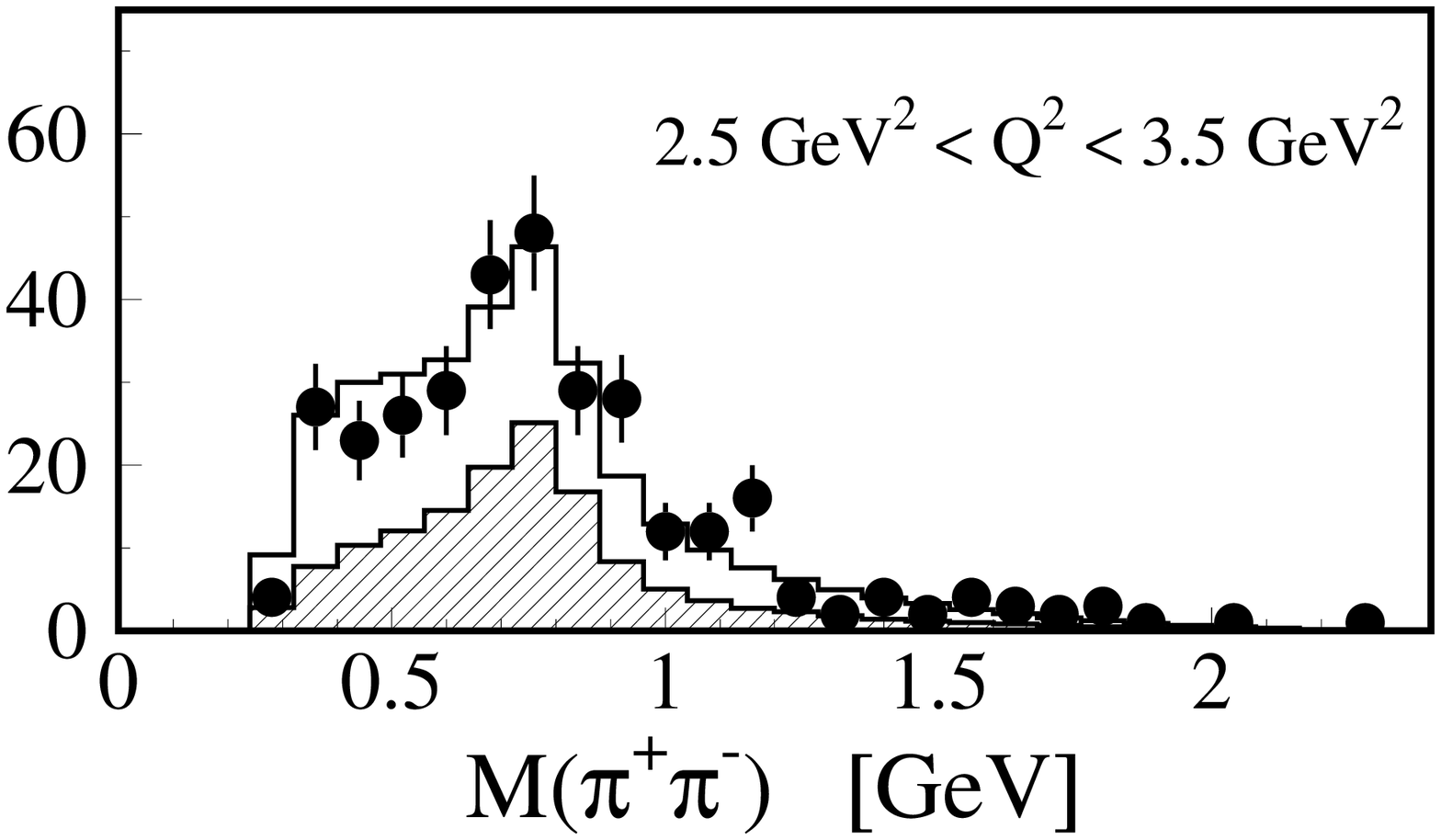,width=0.49\linewidth}}
{\epsfig{file=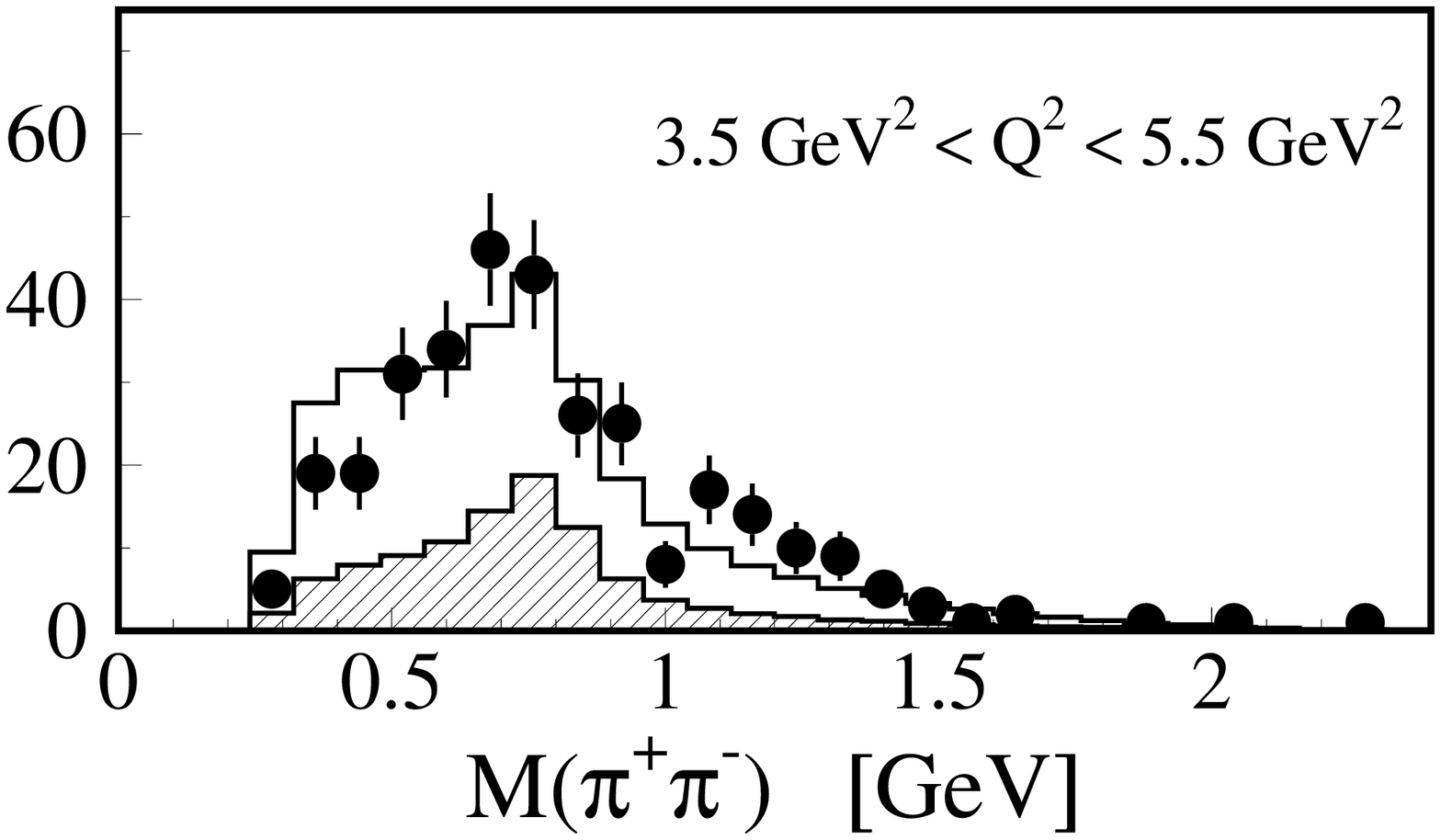,width=0.49\linewidth}}
\vskip -0.5cm
{\epsfig{file=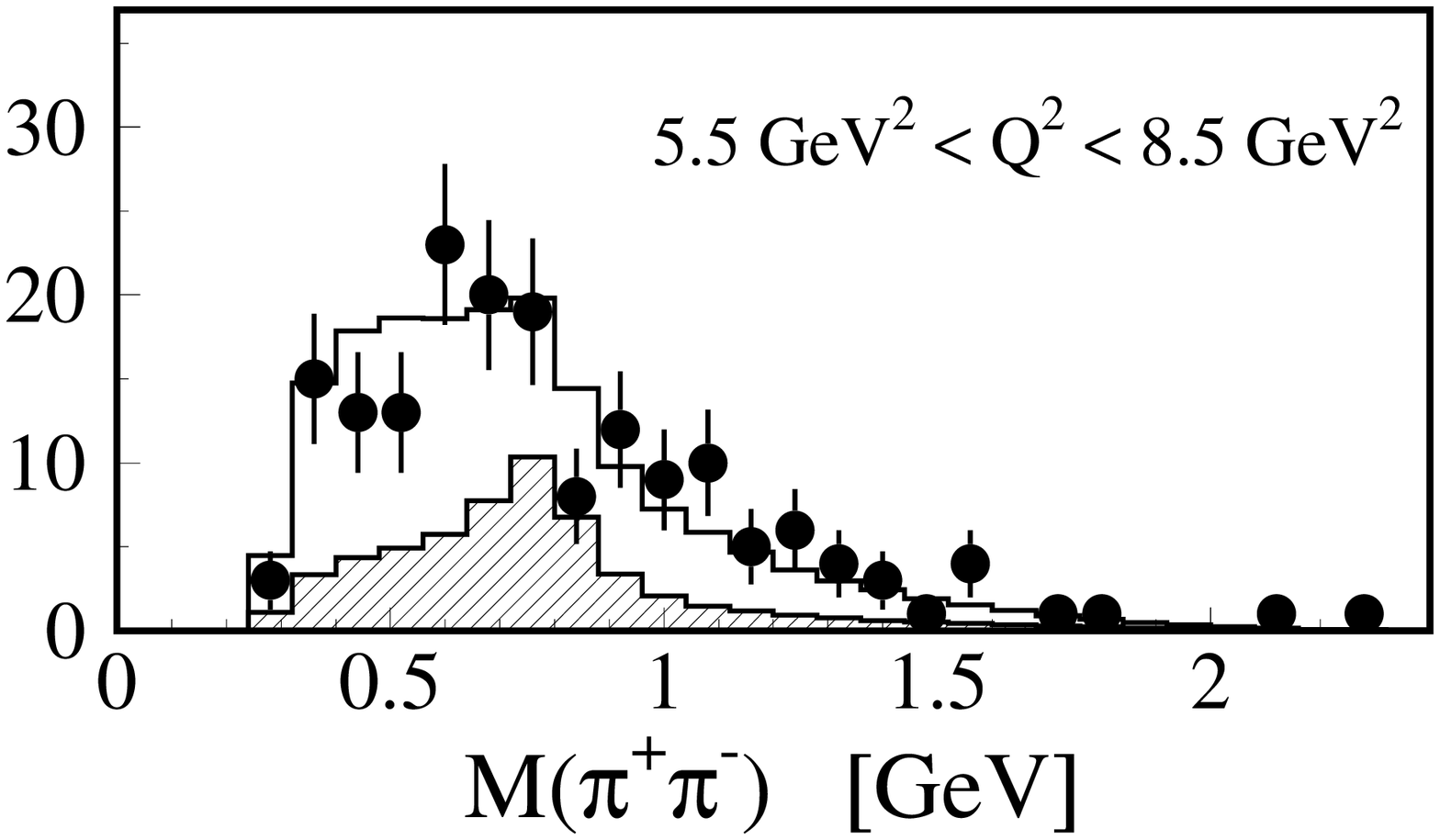,width=0.49\linewidth}}
{\epsfig{file=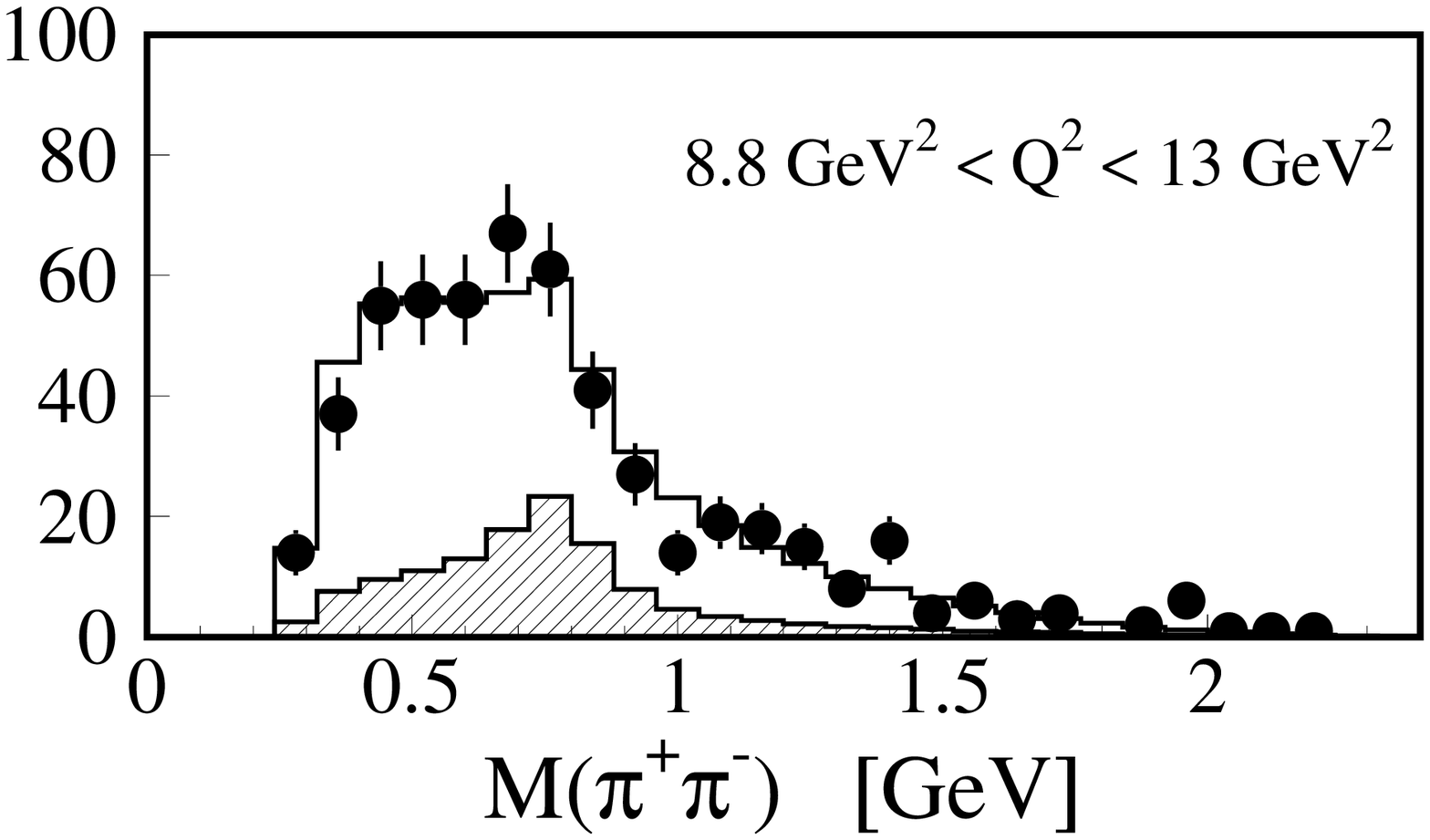,width=0.49\linewidth}}
\vskip -0.5cm
{\epsfig{file=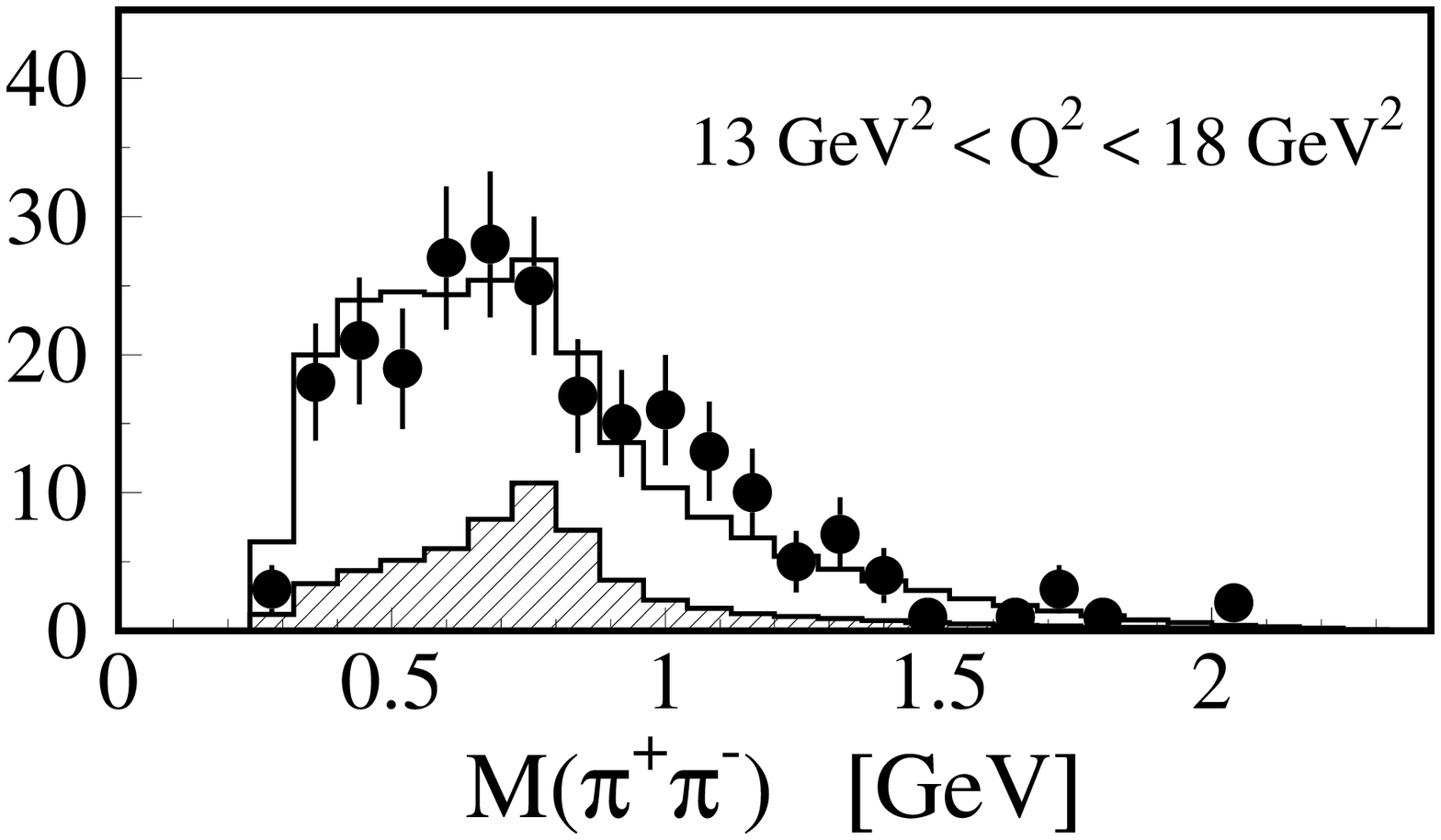,width=0.49\linewidth}}
{\epsfig{file=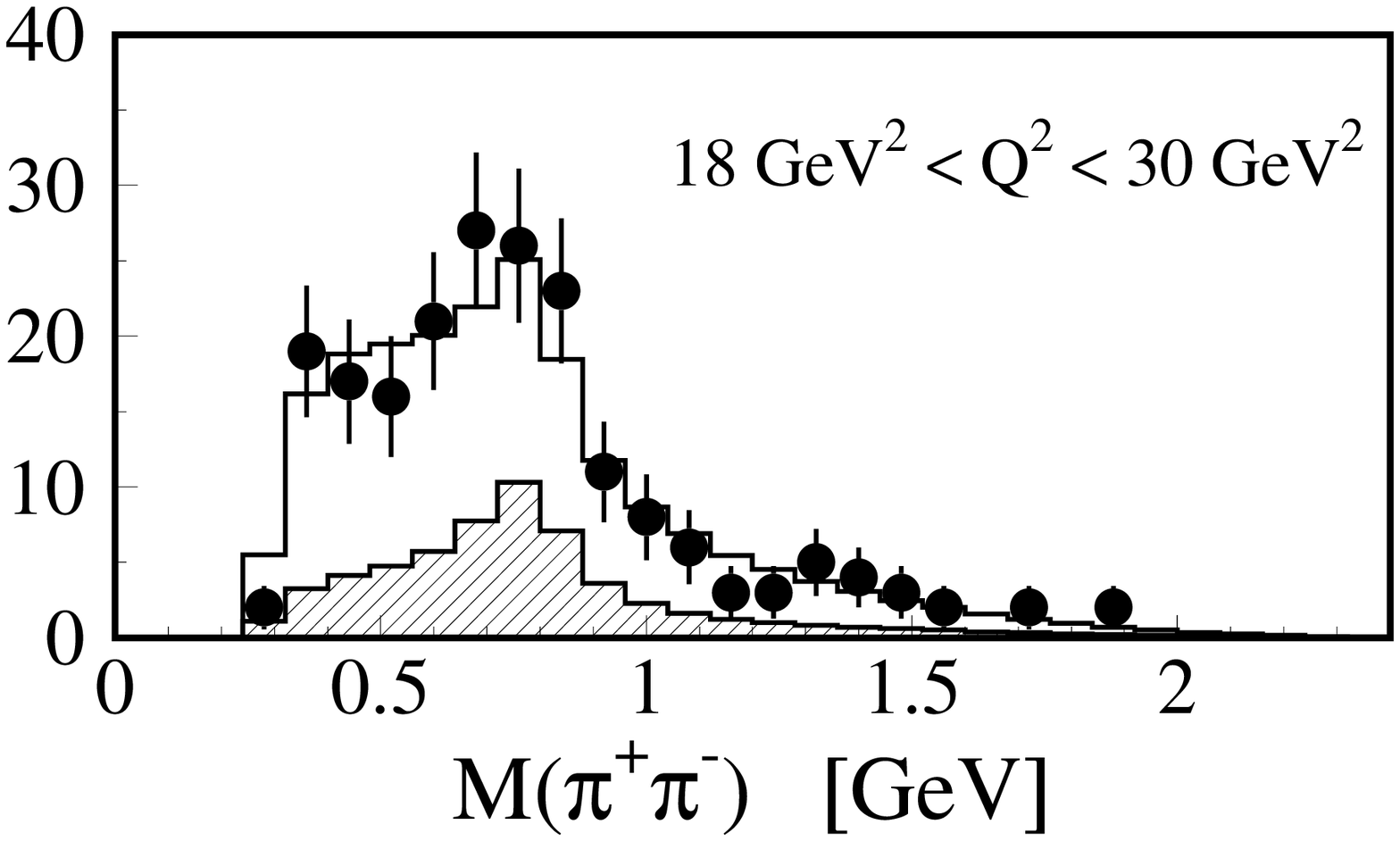,width=0.49\linewidth}}
\end{minipage}

\begin{figure}[h]
\caption[]{Mass distributions of $\pipi$ combinations (four entries per event)
           for events  with $1.1 \GeV < \mgg < 3 \GeV$
           in the fitted $\q$ intervals. 
           The points represent the data, the hatched area  the $\ro\ro$ component,
	   and  the open area   the sum of the
            $\ro \pipi$ and $\pipi\pipi$ (non-resonant) components.
          The fraction of the different components are determined by the fit and the
	  total normalisation is to the number of the events.}
\label{fig:composq2}
\end{figure}

\clearpage


\vfil
\clearpage

\clearpage


\vfil

\clearpage
\vfil

\end{document}